\documentclass[showpacs,amsmath,amssymb,onecolumn,prx,notitlepage,superscriptaddress]{revtex4-2}

\usepackage[dvips]{graphicx} 
\usepackage{amsfonts,amscd,amsmath,amsthm}
\usepackage{mathrsfs}
\usepackage{enumerate}
\usepackage{epsfig}
\usepackage{subcaption}
\usepackage{xcolor}
\usepackage[colorlinks = true]{hyperref}
\usepackage{physics}
\usepackage{epstopdf}
\usepackage{framed}
\usepackage{multirow}
\usepackage{color}
\usepackage{comment}
\usepackage[ruled,vlined]{algorithm2e}
\usepackage[most]{tcolorbox}
\usepackage{pgfplots}
\usepackage{svg}
\usepackage{booktabs}
\usepackage{tabularx}
\usepackage{makecell}

\graphicspath{{./figure/}}

\usepackage{tikz}
\usetikzlibrary{tikzmark, calc, fit, positioning}
\usetikzlibrary{shapes,arrows}
\usetikzlibrary{quantikz}
\pgfplotsset{compat=1.18}

\newtheorem{theorem}{Theorem}
\newtheorem{lemma}{Lemma}

\newtheorem{definition}{Definition}

\newtcolorbox[auto counter]{mybox}[2][]{
	enhanced,
	breakable,
	colback=blue!5!white,
	colframe=blue!75!black,
	fonttitle=\bfseries,
	title=Box \thetcbcounter: #2,#1
}

\newcommand{\id}{\mathbb{I}}
\newcommand{\supp}{\mathrm{supp}}

\renewcommand{\tr}[1]{\mathrm{tr}\left( #1 \right)}

\begin{document}
\preprint{APS/123-QED}

\title{Bypassing the protection-sensitivity incompatibility in quantum-error-corrected metrology via asymmetric codes}

\author{Junjie Chen}
\affiliation{Center for Quantum Information, Institute for Interdisciplinary Information Sciences, Tsinghua University, Beijing, 100084 China}
\author{Rui Luo}
\affiliation{Center for Quantum Information, Institute for Interdisciplinary Information Sciences, Tsinghua University, Beijing, 100084 China}
\author{Zhenyu Du}
\affiliation{Center for Quantum Information, Institute for Interdisciplinary Information Sciences, Tsinghua University, Beijing, 100084 China}
\author{Yuxuan Yan}
\affiliation{Center for Quantum Information, Institute for Interdisciplinary Information Sciences, Tsinghua University, Beijing, 100084 China}
\author{You Zhou}
\email{you\_zhou@fudan.edu.cn}
\affiliation{Key Laboratory for Information Science of Electromagnetic Waves (Ministry of Education), Fudan University, Shanghai 200433, China}
\author{Xiongfeng Ma}
\email{xma@tsinghua.edu.cn}
\affiliation{Center for Quantum Information, Institute for Interdisciplinary Information Sciences, Tsinghua University, Beijing, 100084 China}

\begin{abstract}
Quantum metrology surpasses the classical precision limit by encoding signals in a probe state such that signal contributions are indistinguishable and their phases accumulate coherently as a collective response. In realistic settings, scalable quantum metrology requires quantum error correction to protect the probe against noise. However, quantum error correction relies on syndrome information that distinguishes errors for identification and correction. Because signals and errors act on the same physical degrees of freedom, there is a structural incompatibility between signal-sensitivity and noise-protection in quantum-error-corrected metrology. We quantify this incompatibility by establishing trade-offs between code distance and the quantum Fisher information of code states for non-degenerate codes, quantum low-density parity-check codes, and generalized Shor codes. We bypass this limitation with asymmetric quantum error correction, in which protection is relaxed along the signal direction while being maintained in complementary directions. We construct such codes for local sensing Hamiltonians, restoring Heisenberg-limited precision while retaining a growing distance and hence protection against local perturbations in complementary directions. Strongly asymmetric quantum low-density parity-check and concatenated asymmetric constructions make the framework sparse, scalable, and continuously tunable. The associated probe states can be prepared by constant-depth adaptive circuits with optimal resource scaling.

\end{abstract}

\maketitle



Precise measurement and sensing are central to science and technology, with applications ranging from timekeeping and gravimetry to gravitational-wave detection~\cite{Roslund2024clock,Stray2022gravity,Yu2020LIGO}. The achievable precision is therefore a key figure of merit, and in conventional approaches it is fundamentally limited by statistical fluctuations, scaling inversely with the square root of the number of measurements, $O(1/\sqrt{n})$, a phenomenon known as the standard quantum limit. Quantum metrology exploits quantum coherence and entanglement to make the response of many probes add coherently, approaching sensitivities that scale inversely with the number of measurements, $O(1/n)$, known as the Heisenberg limit~\cite{Giovannetti2004Quantum,Giovannetti2006metrology}, thereby enabling substantial improvements in sensing and timekeeping across a wide range of physical platforms~\cite{Yu2020LIGO,McCormick2019Quantum,Franke2023Quantum,Malia2022Distributed}. In a typical metrological protocol, a quantum probe state undergoes a unitary evolution generated by a Hamiltonian that encodes the unknown parameter, followed by a measurement to estimate its value. The central goal is therefore to design probe states and measurement protocols that allow the signal to accumulate coherently across the entire system. This translates directly into improved precision scaling, making quantum metrology a particularly transparent route toward practical quantum advantage.


In realistic implementations, however, the advantages promised by quantum metrology are fragile in the presence of noise, such as decoherence and control imperfections. Such noise inevitably accumulates in large systems and can rapidly destroy the scaling advantage offered by quantum resources, reducing the achievable precision back to the standard quantum limit~\cite{Huelga1997Improvement,Escher2011General,Demkowicz2012elusive}. Overcoming this challenge requires quantum error correction (QEC) to protect metrological protocols against noise. By encoding quantum states into protected subspaces, QEC enables robust storage and manipulation of quantum information even in the presence of noise~\cite{Terhel2015QEC}. This idea has motivated a substantial body of work on QEC-assisted metrology, showing that error correction can stabilize sensing protocols and, in suitable settings, recover Heisenberg-limited scaling~\cite{Kessler2014quantum,Dur2014improved,zhou2018achieving,Mann2025Quantum}. However, most existing approaches start from a specified noise model and ask whether that noise can be corrected without destroying the signal. The constructions depend on particular signal-noise relations, including perpendicular-noise, Hamiltonian-not-in-Lindblad-span and commuting-noise settings, often together with fast control, noiseless ancillas, or constructions tailored to specific small-scale scenarios~\cite{Demkowicz2017Adaptive,Layden2019Ancilla,Rojkov2022Bias,Kwon2026virtual}. At this point, the general role of QEC in metrology remains unresolved: can QEC protect a sensor without suppressing the very signal it is meant to amplify? This structural issue is central to a general theory of scalable QEC-assisted metrology.

\begin{figure}[htbp]
\centering

\begin{subfigure}[t]{\textwidth}
\centering
\includegraphics[width=\linewidth]{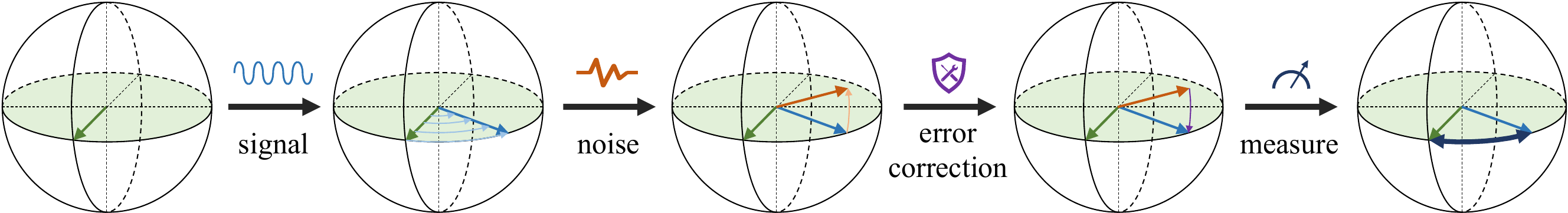}
\caption{}
\label{fig:procedure}
\end{subfigure}

\begin{subfigure}[t]{\textwidth}
\centering
\includegraphics[width=0.85\linewidth]{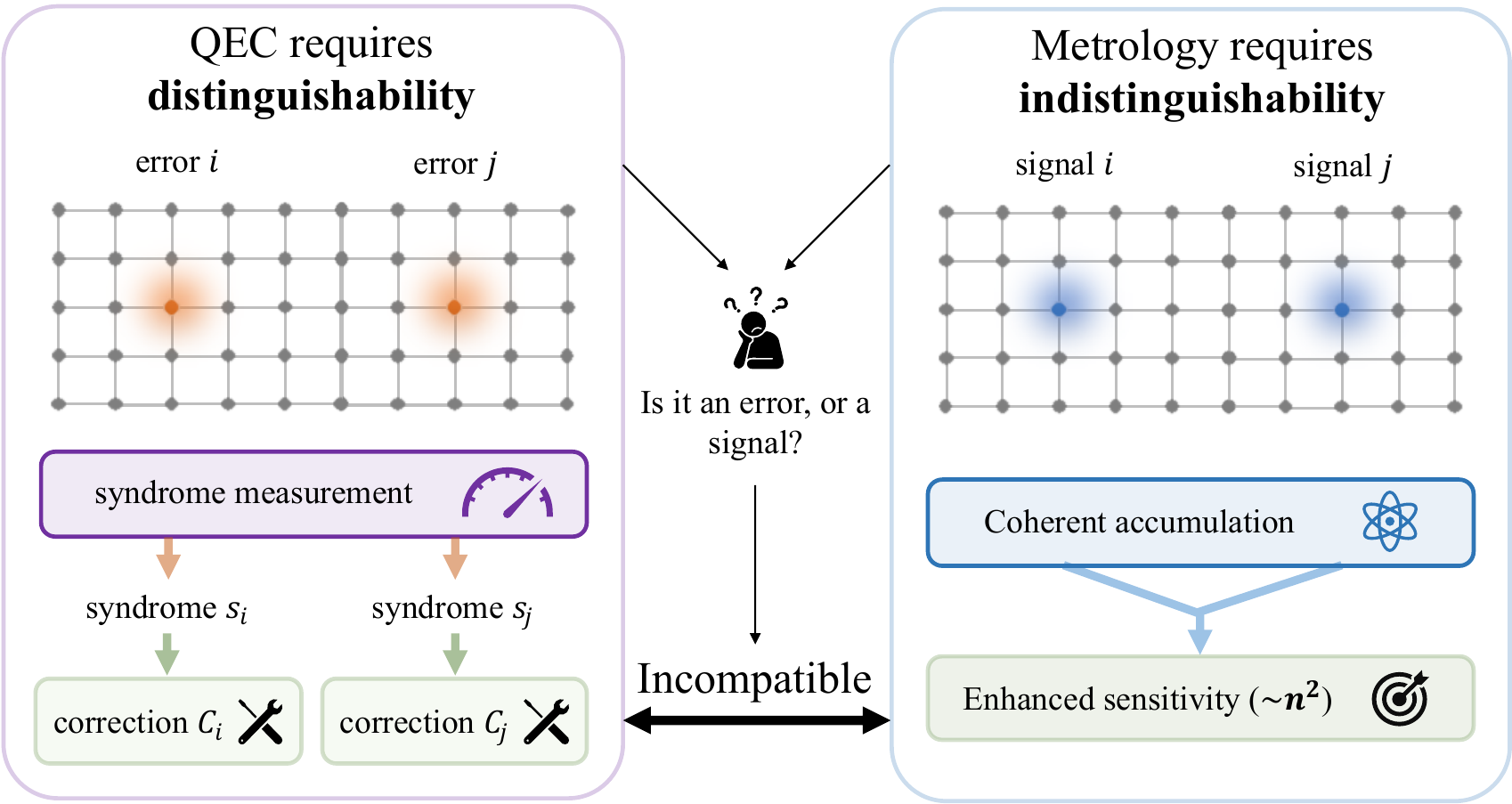}
\caption{}
\label{fig:incompatible}
\end{subfigure}

\caption{
\textbf{(a)} A kind of QEC-assisted sensing protocol. After preparing a probe code state, the signal Hamiltonian imprints a parameter-dependent phase on it. If the process is disturbed by noise, we can use QEC to remove it before the final measurement to get the information about the signal.
\textbf{(b)} Incompatibility between protection and sensitivity. QEC requires local errors at different positions, or in different error directions, to be distinguishable through syndrome information so that they can be identified and corrected. By contrast, quantum-enhanced metrology requires local signal contributions to be indistinguishable within the probe state, so that they add coherently and generate a macroscopic response, enabling Heisenberg-limited sensitivity. Since a local signal and a local error act through the same physical degrees of freedom, symmetric QEC, which suppresses all local perturbations, also suppresses the coherent accumulation required for enhanced metrological sensitivity. This leads to the incompatibility between protection and sensitivity.
}
\label{fig:introduction}
\end{figure}

We answer this question by showing that the requirements of quantum metrology and quantum error correction are intrinsically incompatible. As illustrated in Fig.~\ref{fig:incompatible}, quantum metrology needs local signal contributions to remain indistinguishable within the probe state, so that they add coherently. By contrast, QEC protects against local noise by extracting syndrome information that distinguishes local errors. Therefore, symmetrically protected quantum-error-corrected metrology faces a fundamental structural incompatibility between signal indistinguishability and error distinguishability, since local signal and local errors act through the same physical degrees of freedom. We make this intuition quantitative through a protection-sensitivity trade-off, where protection is measured by the code distance $d$ and sensitivity by the quantum Fisher information (QFI) $\mathcal{F}$, $d\cdot\mathcal{F}=O(n^2)$. In QFI language, $\mathcal{F}=O(n)$ corresponds to standard-quantum-limit precision, whereas $\mathcal{F}=\Theta(n^2)$ corresponds to Heisenberg-limited precision. For nondegenerate codes~\cite{bjork2009efficiency}, once the distance exceeds the locality scale of the sensing Hamiltonian, every code state has $\mathcal{F}=O(n)$ under local Hamiltonian. We then show that degeneracy alone is not enough to remove the incompatibility. For quantum low-density parity-check (QLDPC) codes~\cite{Kitaev2003Fault,Panteleev2022GoodLDPC} with growing distance, the sparse-check structure also enforces $\mathcal{F}=O(n)$. If the QLDPC constraint is relaxed, Shor-type degeneracy~\cite{Bacon2006Operator} can correlate local signal terms and achieve both nonconstant distance and superlinear QFI. Even then, the enhancement is still limited by the trade-off $d\cdot\mathcal{F}=O(n^2)$. Thus, degeneracy opens intermediate regimes between the standard quantum and Heisenberg limits, but does not remove the cost of protection.

We then bypass this incompatibility by replacing symmetric protection with directional protection. The key insight is that a sensor need not protect all local directions equally: the signal direction should remain accessible, while complementary directions should remain correctable. We formalize this principle as asymmetric QEC and construct such codes for local Hamiltonians. In these codes, local signal terms are deliberately left indistinguishable within the code space and therefore add coherently, yielding Heisenberg-limited QFI, $\mathcal{F}=\Theta(n^2)$. By contrast, local perturbations in non-signal directions remain distinguishable and correctable with growing effective distance. Asymmetric QEC therefore preserves exactly the response that metrology needs while retaining error protection in the directions that would otherwise carry noise.

We further develop scalable variants that make this principle both tunable and implementable. Strongly asymmetric quantum low-density parity-check codes give sparse-check constructions with constant effective distance along the signal direction, growing distance in all the other complementary directions, and Heisenberg-limited QFI. Concatenated asymmetric codes, inspired by generalized Shor codes, make the balance continuously tunable: by adjusting the residual protection along the signal direction, they interpolate between highly sensitive probes and more strongly protected probes, while keeping complementary directions better protected than symmetric codes with the same signal sensitivity would allow. Finally, we show that these probe states can be prepared by constant-depth adaptive circuits, with resources controlled by the total stabilizer-check weight and with optimal scaling for the strongly asymmetric QLDPC constructions.

\section{Results}

\subsection{Settings}

The precision of measurement and sensing is fundamentally limited by statistical fluctuations. Quantum metrology provides a framework to characterize and overcome these limits by exploiting quantum coherence and entanglement. In this framework, the goal is to estimate an unknown parameter $\theta$ encoded in a quantum state through a unitary evolution $\rho(\theta) = e^{-i\theta \hat{H}} \rho e^{i\theta \hat{H}}$ generated by a Hamiltonian $\hat{H}$. Given such a quantum state $\rho(\theta)$, measurements are performed to construct an unbiased estimator $\hat{\theta}$ for $\theta$. The achievable precision is bounded by the quantum Cramér–Rao inequality~\cite{Braunstein1994statistical,Paris2009quantum},
\begin{equation}\label{eq:QCRbound}
(\Delta \hat{\theta})^2 \geq \frac{1}{\mathcal{F}(\rho(\theta))},
\end{equation}
where $\mathcal{F}(\rho(\theta))$ is the QFI, which quantifies the maximum information about the parameter that can be extracted from the state. For $N$ independent copies of $\rho_{\theta}$, the variance can be reduced by $N$ times. Based on Eq.~\eqref{eq:QCRbound}, the achievable precision is directly determined by how the QFI scales with the system size. In particular, if the QFI scales linearly as $\mathcal{F}=O(n)$, then the precision scales as $\Delta \hat{\theta}=O(1/\sqrt{n})$, known as the standard quantum limit. By contrast, if the QFI scales quadratically as $\mathcal{F}=O(n^2)$, then the precision scales as $\Delta \hat{\theta}=O(1/n)$, known as the Heisenberg limit. Classical sensing strategies cannot surpass the standard quantum limit, whereas quantum metrology can, in principle, achieve the Heisenberg limit, thereby demonstrating a quantum advantage in parameter estimation. Since $\rho(\theta)$ is determined by the probe state $\rho$ and the Hamiltonian $\hat{H}$, the achievable precision depends crucially on both. When the probe state is pure, the QFI is given by
\begin{equation}\label{eq:QFIdecom}
\mathcal{F}(\ket{\psi},\hat{H})=\bra{\psi}\hat{H}^2\ket{\psi}-\left|\bra{\psi}\hat{H}\ket{\psi}\right|^2.
\end{equation}

In realistic sensing scenarios, the Hamiltonians that encode the unknown parameter are typically local, reflecting the structure of physical interactions in many-body quantum systems. We therefore focus on $K$-local Hamiltonians, which capture a broad and physically relevant class of metrological settings. First, without loss of generality, we assume that $\hat{H}$ is traceless, since any identity component of the Hamiltonian contributes only a global phase to the evolution and therefore does not affect the QFI. Specifically, we consider a $K$-local traceless Hamiltonian of the form
\begin{equation}\label{eq:Hdecom}
\hat{H} = \sum_{j=1}^{m} \hat{H}_j ,
\end{equation}
with $K=O(1)$ where each traceless term $\hat{H}_j$ acts nontrivially on at most $K$ qubits and each qubit participates in at most $K$ terms in this decomposition. 
In addition, the operator norm of $\hat{H}$ scales as $\|\hat{H}\|_{\infty}=\Theta(n)$ and each term is bounded as $\|\hat{H}_j\|_{\infty} \leq 1$. These conditions ensure that the Hamiltonian represents a physically meaningful many-body interaction, in which each local term has bounded strength and each qubit contributes only a finite amount to the total signal. Under these assumptions, the number of terms scales linearly with the system size, $m=\Theta(n)$. We keep the extensive scaling as $\|\hat{H}\|_{\infty}=\Theta(n)$ in particular to guarantee that any enhancement in precision arises from collective quantum effects across the system, rather than from artificially amplifying the Hamiltonian, so that the scaling of the QFI faithfully reflects the intrinsic metrological advantage. Our goal is to understand how the QFI $\mathcal{F}(\ket{\psi},\hat{H})$ can scale with the system size $n$ under these physically motivated constraints, and in particular how quantum error-correcting code states influence the achievable metrological scaling.

\subsection{Protection-sensitivity incompatibility in QEC metrology}

QEC code states are protected against noise through highly structured many-body correlations and thus appear as natural probe states for noise-resilient quantum metrology. However, we show that this natural strategy contains an intrinsic incompatibility: the same mechanism that protects encoded information from local errors can also suppress the response to a local signal. Thus, there should be an incompatibility between error protection and signal sensitivity, captured by two figures of merit. The code distance $d$ characterizes the strength of local-error protection, while the QFI $\mathcal{F}(\ket{\psi},\hat{H})$ characterizes the sensitivity of a probe state to the signal Hamiltonian. Heisenberg-limited metrology requires $\mathcal{F}(\ket{\psi},\hat{H})=\Theta(n^2)$, whereas scalable QEC requires a distance that grows with the system size $n$. For codes with conventional symmetric local-error protection, these two requirements cannot be simultaneously maximized. The resulting trade-off can be stated informally as follows.

\begin{theorem}[Protection-sensitivity incompatibility (informal)]\label{theo:tradeoff}
For principal families of $[[n,k,d]]$ QEC codes, including non-degenerate codes, QLDPC codes, and generalized Shor codes, any code state $\ket{\psi}$ satisfies the following trade-off between the code distance $d$ and QFI $\mathcal{F}$,
\begin{equation}\label{eq:tradeoff}
d\cdot\mathcal{F}(\ket{\psi},\hat{H})=O(n^2),
\end{equation}
under any local Hamiltonian $\hat{H}$ defined in Eq.~\eqref{eq:Hdecom}.
\end{theorem}

To illustrate the trade-off more intuitively, we have the following table and Fig.~\ref{fig:tradeoff}.

\begin{table}[htbp]
\centering
\small
\setlength{\tabcolsep}{10pt}
\caption{Protection-sensitivity trade-off in typical QEC codes.}
\label{table:tradeoff}
\begin{tabular}{@{}cccc@{}}
\toprule
Code family
& Distance $d$
& QFI $\mathcal{F}$
& Trade-off \\
\midrule
Non-degenerate codes
& $2K+1\leq d\leq n$
& $O(n)$
& $d\cdot\mathcal{F}=O(n^2)$ \\
\addlinespace
QLDPC codes
& $2K+1\leq d\leq n$
& $O(n)$
& $d\cdot\mathcal{F}=O(n^2)$ \\
\addlinespace
Generalized Shor codes
& $2K+1 \leq d \leq \sqrt{n}$
& $O(n^2/d)$
& $d\cdot\mathcal{F}=O(n^2)$ \\
\bottomrule
\end{tabular}
\end{table}

We first illustrate the trade-off in non-degenerate codes, where distinct correctable errors have distinct syndromes and map the code space to mutually orthogonal error subspaces~\cite{Knill1997KLcondition}. For the $K$-local Hamiltonian $\hat{H}=\sum_{j=1}^{m}\hat{H}_j$ in Eq.~\eqref{eq:Hdecom}, once the distance satisfies $d\ge 2K+1$, each local signal term is itself correctable, and any code state obeys $\mathcal{F}(\ket{\psi},\hat{H})\le mK^2$. Since $m=\Theta(n)$ and $K=O(1)$, the QFI is at most linear in the system size, as shown in the second row in Table~\ref{table:tradeoff}. Thus, for any non-degenerate code family with non-constant distance, increasing protection enforces standard-quantum-limit scaling rather than Heisenberg scaling.

The reason is that error correction and metrological enhancement require opposite behavior from local perturbations. As illustrated in Fig.~\ref{fig:incompatible}, QEC requires local errors to be distinguishable through syndrome information, whereas Heisenberg-limited metrology requires local signal terms to remain indistinguishable within the probe state so that their phases add coherently. In a non-degenerate code with $d\geq 2K+1$, the Knill-Laflamme conditions eliminate correlations between signal terms supported on disjoint regions: their actions either reduce to scalars on the code space or belong to orthogonal error subspaces. Since the QFI is the variance of $\hat{H}$, it is a sum of two-point correlations between $\hat{H}_j$ and $\hat{H}_{j'}$. Only overlapping local terms can contribute, and bounded locality leaves only $O(mK^2)$ such pairs. A formal statement and proof are given in the Supplementary Information.

We next consider degenerate codes. In this setting, the non-degenerate argument no longer applies: distinct correctable errors need not occupy orthogonal syndrome subspaces and may therefore interfere coherently. This degeneracy could, in principle, allow local signal terms to accumulate and evade the simple correlation-counting argument above.

For quantum low-density parity-check (QLDPC) codes, however, locality restores essentially the same limitation. QLDPC codes are widely studied for scalable quantum information processing due to their local stabilizer structure: each stabilizer acts on a constant number of qubits, and each qubit participates in only a constant number of stabilizers~\cite{Gottesman2014Faulttolerant,Dinur2023GoodQLDPC,Bravyi2024FTmemory,Zhang2025Time}. This sparse stabilizer structure strongly restricts the correlations entering the QFI: each local signal term can be coherently correlated with only a bounded number of nearby local terms. Long-range correlations would require either a low-weight logical operator, which contradicts the growing-distance assumption, or a long-range stabilizer, which contradicts the locality property of QLDPC codes. Consequently, for any $K$-local Hamiltonian $\hat{H}$ in Eq.~\eqref{eq:Hdecom} with $K=O(1)$, any QLDPC code state with $d\geq 2K+1$ satisfies $\mathcal{F}(\ket{\psi},\hat{H})=O(n)$, as shown in the third row in Table~\ref{table:tradeoff}. Thus, even with degeneracy, local stabilizer structure prevents Heisenberg-limited scaling at the physical-qubit level. The formal statement and proof are given in the Supplementary Information.

Degenerate codes without a locality constraint are more subtle. Degeneracy can turn products of local signal operators into stabilizers, allowing signal terms that would be independent in a non-degenerate code to add coherently. Generalized Shor codes~\cite{Bacon2006Operator} provide a useful testbed for this mechanism. They show that degeneracy can enhance sensitivity, but only at the cost of protection: the more local signal terms are allowed to accumulate coherently, the smaller the resulting code distance becomes.

We consider CSS generalized Shor codes constructed from two classical binary linear codes, with the physical qubits arranged into blocks. The $Z$-type stabilizers generated by one classical code act within each block, whereas the $X$-type stabilizers given by the other one couple different blocks. By tuning the two classical codes, such constructions can realize distances ranging from constant values up to $O(\sqrt{n})$. For any $K$-local Hamiltonian in Eq.~\eqref{eq:Hdecom} with $K=O(1)$, we prove that any generalized Shor-code state with distance $d\geq 2K+1$ satisfies $\mathcal{F}(\ket{\psi},\hat{H})=O(n^2/d)$, as shown in the final row in Table~\ref{table:tradeoff}. Thus, generalized Shor codes can evade the linear-QFI limitation of non-degenerate and QLDPC codes, but they still obey the same protection-sensitivity trade-off. The formal definition of generalized Shor codes and the proof of this bound are given in the Supplementary Information.

The generalized Shor-code example also gives intuition for why the trade-off should extend beyond this specific construction. Consider, for instance, a stabilizer code and decompose the signal Hamiltonian into local Pauli components. The QFI is controlled by connected correlations between these Pauli terms. Degeneracy can organize local signal terms into coherent clusters: terms within the same cluster have nonzero correlations and add coherently, whereas different clusters must remain uncorrelated, or they would merge into a larger cluster. Enlarging these clusters increases the QFI, with the enhancement determined by the cluster size. At the same time, symmetric local-error protection requires enough stabilizer structure to distinguish and protect the different clusters, so the achievable distance is controlled by how many such independent clusters remain. This is precisely the structure realized by generalized Shor codes, where the block size controls coherent signal accumulation, and the number of blocks controls the distance. This picture suggests that, for QEC codes with symmetric local-error protection, any degeneracy that enhances sensitivity by coherent accumulation must simultaneously reduce the distance, leading generically to the trade-off $d\cdot\mathcal{F}(\ket{\psi},\hat{H})=O(n^2)$.

As shown in Table~\ref{table:tradeoff}, the trade-off is tight at the level of scaling. At one endpoint, an unprotected GHZ probe under a collective local Hamiltonian, such as $\hat{H}=\sum_j Z_j$, has $\mathcal{F}(\ket{\psi},\hat{H})=\Theta(n^2)$; viewed as a trivial distance-one code, it gives $d\cdot\mathcal{F}(\ket{\psi},\hat{H})=\Theta(n^2)$. At the opposite endpoint, non-degenerate codes and QLDPC constructions can realize linear distance, $d=\Theta(n)$, while the local signal terms contribute only incoherently, giving $\mathcal{F}(\ket{\psi},\hat{H})=\Theta(n)$ for local Hamiltonians. These examples also saturate $d\cdot\mathcal{F}(\ket{\psi},\hat{H})=\Theta(n^2)$, but in the regime of strong protection and standard-quantum-limit sensitivity. Generalized Shor codes fill an intermediate part of the same trade-off. Choose a Shor-type construction with $d$ blocks, each of size $n/d$, so that the code distance is $d$ for $d\leq \sqrt{n}$. For the blockwise GHZ code state and a block-aligned Hamiltonian, such as $\hat{H}=\sum_j Z_j$, signal terms add coherently within each block but not between different blocks, giving QFI $\mathcal{F}(\ket{\psi},\hat{H})=\Theta(n^2/d)$. Thus, generalized Shor codes saturate the trade-off, throughout the range from constant distance to $d=\Theta(\sqrt{n})$. The ordinary Shor code sits at the endpoint $d=\Theta(\sqrt{n})$, with $\mathcal{F}=\Theta(n^{3/2})$. This tunable saturation illustrates how degeneracy enhances sensitivity: it permits coherent accumulation inside larger blocks, but increasing the size of such blocks necessarily reduces the number of blocks available for protection, and hence reduces the distance.

We emphasize that our results cover three representative mechanisms by which exact QEC codes protect against local errors: nondegenerate syndrome separation, sparse-check local protection as in QLDPC architectures, and Shor-type degeneracy. This breadth suggests that the resulting trade-off is not an artifact of any particular code construction. We further remark that the trade-off is distinct from previous no-go results based on the relation between signal and noise, such as the Hamiltonian-not-in-Lindblad-span condition~\cite{Demkowicz2017Adaptive,zhou2018achieving}. Those results ask whether a specified noise process can be corrected without erasing the signal. By contrast, the present trade-off does not rely on any noise model: it already applies to noiseless parameter encoding and shows that sufficiently strong local-error protection can itself suppress the coherent accumulation of local signals. The incompatibility, therefore, originates from the error-correcting structure of the code, rather than from an unfavorable relation between a particular noise process and the signal Hamiltonian.

The trade-off should nevertheless be understood as a statement about protection and sensitivity within the same physical degrees of freedom. Otherwise, it can be bypassed in a trivial way. For example, one may take a product state consisting of a high-distance stabilizer-code state on half of the qubits and a GHZ state on the remaining qubits, and let the signal Hamiltonian act as $\hat{H}=\sum_j Z_j$. The GHZ factor alone gives $\mathcal{F}(\ket{\psi},\hat{H})=\Theta(n^2)$, while the unused code factor can carry non-constant distance. This construction, however, is not a protected metrological code: the qubits responsible for the Heisenberg-limited response are not protected by the high-distance code. Indeed, protecting the GHZ factor itself would require checking its stabilizers during sensing; the stabilizer that certifies the GHZ coherence is precisely conjugate to the accumulated phase, so such checks would erase the signal, whereas omitting them leaves the GHZ subsystem unprotected. This observation suggests that protection and sensitivity should not be separated between different subsystems, but rather between different error directions on the same physical degrees of freedom, which naturally leads to asymmetric QEC in the next section.

\begin{figure}
\centering

\begin{subfigure}[t]{0.47\textwidth}
\definecolor{Allowed}{HTML}{E2F0D9}
\definecolor{Forbidden}{HTML}{F5D9D2}
\definecolor{Guide}{HTML}{222222}
\definecolor{Accent}{HTML}{2357A6}
\centering
\raisebox{0mm}{
\begin{tikzpicture}[
    x=0.9cm,y=0.9cm,
    >=Latex,
    axis/.style={line width=1.0pt,->},
    guide/.style={densely dashed,line width=0.55pt,draw=Guide!75},
    boundary/.style={line width=1.55pt,draw=black,line cap=round},
    tick/.style={line width=0.8pt,draw=black},
    label/.style={align=center},
    smalllabel/.style={align=center},
    domain=1:6, samples=100
]
\def\xmax{6.15}
\def\ymax{6.15}
\def\xOne{1}
\def\xSqrt{2.44948974278}
\def\xN{6}
\def\ySQL{1}
\def\yMid{2.44948974278}
\def\yHL{6}
\path[fill=Allowed]
    (0,0) -- (0,\yHL) -- (\xOne,\yHL) -- plot (\x, {6/\x}) -- (\xN,\ySQL) -- (\xN,0) -- cycle;
\draw[guide] (0,\yHL) -- (\xmax,\yHL);
\draw[guide] (0,\ySQL) -- (\xmax,\ySQL);
\draw[guide] (0,\yMid) -- (\xSqrt,\yMid);
\draw[guide] (\xOne,0) -- (\xOne,\ymax);
\draw[guide] (\xSqrt,0) -- (\xSqrt,\yMid);
\draw[guide] (\xN,0) -- (\xN,\ymax);
\draw[boundary] plot (\x, {6/\x});
\draw[axis] (0,0) -- (\xmax+0.45,0);
\draw[axis] (0,0) -- (0,\ymax+0.35);
\node[label,below=22pt] at (0.55*\xmax,0) {Protection $d$};
\node[label,rotate=90] at (-1.90,0.53*\ymax) {Sensitivity $\mathcal{F}$};
\foreach \x/\txt in {\xOne/{\(\Theta(1)\)},\xSqrt/{\(\Theta(\sqrt n)\)},\xN/{\(\Theta(n)\)}}
{\draw[tick] (\x,0) -- ++(0,-0.10);\node[label,below=5pt] at (\x,0) {\txt};}
\foreach \y/\txt in {\yHL/{\(\Theta(n^2)\)},\yMid/{\(\Theta(n^{3/2})\)},\ySQL/{\(\Theta(n)\)}}
{\draw[tick] (0,\y) -- ++(-0.10,0);\node[label,left=7pt] at (0,\y) {\txt};}
\node[smalllabel,text opacity=1,inner sep=1.5pt,rounded corners=1pt] at (4.8,2) {$d\cdot\mathcal F=O(n^2)$};
\filldraw[fill=black,draw=black,line width=0.7pt] (\xSqrt,\yMid) circle (2.0pt);
\filldraw[fill=black,draw=black] (\xOne,\yHL) circle (1.7pt);
\filldraw[fill=black,draw=black] (\xN,\ySQL) circle (1.7pt);
\end{tikzpicture}
}
\caption{}
\label{fig:tradeoff}
\end{subfigure}
\hspace{0.02\textwidth}
\begin{subfigure}[t]{0.49\textwidth}
\centering

\definecolor{Guide}{HTML}{222222}
\definecolor{QFIHigh}{HTML}{3B0F70}   
\definecolor{QFIMidA}{HTML}{0B6E4F}   
\definecolor{QFIMidB}{HTML}{8CC63E}   
\definecolor{QFILow}{HTML}{F6E84A}    
\definecolor{Construct}{HTML}{F03A2E} 
\pgfdeclarehorizontalshading{qfiInverse}{100bp}{%
  color(0bp)=(QFIHigh);
  color(18bp)=(QFIHigh);
  color(42bp)=(QFIMidA);
  color(72bp)=(QFIMidB);
  color(100bp)=(QFILow)}
\pgfdeclareverticalshading{qfiLegend}{100bp}{%
  color(0bp)=(QFILow);
  color(28bp)=(QFIMidB);
  color(58bp)=(QFIMidA);
  color(100bp)=(QFIHigh)}

\begin{tikzpicture}[
    x=0.9cm,y=0.9cm,
    >=Latex,
    axis/.style={line width=1.05pt,->},
    guide/.style={densely dashed,line width=0.55pt,draw=Guide!75},
    mainline/.style={line width=1.25pt,draw=black,line cap=round},
    construct/.style={line width=1.6pt,draw=Construct,line cap=round},
    tick/.style={line width=0.8pt,draw=black},
    label/.style={align=center},
    smalllabel/.style={align=center},
    domain=0:1.64575131106, samples=100
]
\def\xN{6}
\def\yN{6}
\def\xMid{1.64575131106}
\def\yMid{1.64575131106}
\def\bandw{0.24}

\begin{scope}[shift={(7,2.2)}]
    \shade[shading=qfiLegend,draw=black,line width=0.8pt,rounded corners=1pt]
        (0,0) rectangle (0.95,2.95);
    \node[label,anchor=west] at (1,2.75) {\(\Theta(n^2)\)};
    \node[label,anchor=west] at (1,0.18) {\(\Theta(n)\)};
    \node[label] at (0.48,-0.58) {QFI};
\end{scope}

\begin{scope}[shift={(0,0)}]
\begin{scope}
    \clip (0,0) -- (0,\yN) -- (\xN,\yN) -- cycle;
    \shade[shading=qfiInverse] (0,0) rectangle (\xN,\yN);
\end{scope}
\draw[guide] (0,\yN) -- (6.65,\yN);
\draw[guide] (\xN,0) -- (\xN,6.65);
\draw[guide] (0,\yMid) -- (\xMid,\yMid);
\draw[guide] (\xMid,0) -- (\xMid,\yMid);
\draw[mainline] (0,0) -- (\xN,\yN);
\filldraw[fill=none,draw=red, thick]
    (0,0) -- (0,\yN) -- plot (\x, {7/(\x+1)-1}) -- (\xMid,\yMid) -- cycle;
\draw[axis] (0,0) -- (6.9,0) node[below right=2pt] {$d_{\mathrm{sig}}$};
\draw[axis] (0,0) -- (0,6.9) node[left=2pt] {$d_{\mathrm{comp}}$};
\foreach \x/\txt in {0/{\(\Theta(1)\)},\xMid/{\(\Theta(\sqrt n)\)},\xN/{\(\Theta(n)\)}}{
    \draw[tick] (\x,0) -- ++(0,-0.10);
    \node[label,below=4pt] at (\x,0) {\txt};
}
\foreach \y/\txt in {\yMid/{\(\Theta(\sqrt n)\)},\yN/{\(\Theta(n)\)}}{
    \draw[tick] (0,\y) -- ++(-0.10,0);
    \node[label,left=4pt] at (0,\y) {\txt};
}
\node[label] at (3,-1.05)
    {Asymmetric QEC, $d_{\mathrm{comp}}\gtrsim d_{\mathrm{sig}}$};
\end{scope}
\end{tikzpicture}
\caption{}
\label{fig:asymmetricproperty}
\end{subfigure}
\caption{Summary of our results.
\textbf{(a)} For symmetrically protected codes including non-degenerate codes, QLDPC codes, and generalized Shor codes, the code distance $d$, which quantifies local-error protection, constrains the quantum Fisher information (QFI) $\mathcal{F}$, which quantifies metrological sensitivity. The accessible region is bounded by the scaling trade-off $d\cdot\mathcal{F}=O(n^2)$, which implies Heisenberg-limited sensitivity $\mathcal{F}=\Theta(n^2)$ is compatible only with constant protection, whereas growing distance forces lower QFI. The point $(\Theta(1),\Theta(n^2))$ can be reached by unprotected states such as GHZ states; The point $(\Theta(n),\Theta(n))$ can be reached by non-degenerate codes or QLDPC codes with distance $\Theta(n)$; The point $(\Theta(\sqrt{n}),\Theta(n^{3/2}))$ can be reached by Shor codes.
\textbf{(b)} Asymmetric QEC bypasses the incompatibility by separating the signal-direction distance $d_{\mathrm{sig}}$ from the complementary-direction distance $d_{\mathrm{comp}}$. Symmetric QEC lies around the black line, $d_{\mathrm{sig}}\sim d_{\mathrm{comp}}$, where increasing protection also suppresses the signal. Asymmetric QEC can keep $d_{\mathrm{sig}}$ small so that local signal terms remain accessible, while $d_{\mathrm{comp}}$ grows to protect against complementary local errors, enabling Heisenberg-limited QFI together with directional protection. The area circled in red is where we can provide the corresponding construction for any local Hamiltonian defined in Eq.~\eqref{eq:Hdecom}.
}
\label{fig:results}
\end{figure}

\subsection{Asymmetric quantum error correction for quantum metrology}\label{sec:construction}

The trade-off in the last section should not be interpreted as a failure of QEC-assisted metrology in general, but a consequence of symmetric local-error protection. Conventional QEC codes protect symmetrically against local perturbations in all logical directions. This not only makes local errors correctable, since low-weight operators cannot distinguish states within the code space, but also suppresses the coherent logical action needed for Heisenberg-limited sensitivity, as a local metrological signal is generated by the same type of low-weight operators. Degeneracy can partially soften this constraint, but increasing the relevant distance still reduces the achievable sensitivity.

\begin{figure}
\centering
\begin{subfigure}[t]{0.48\textwidth}
\centering
\includegraphics[width=\linewidth]{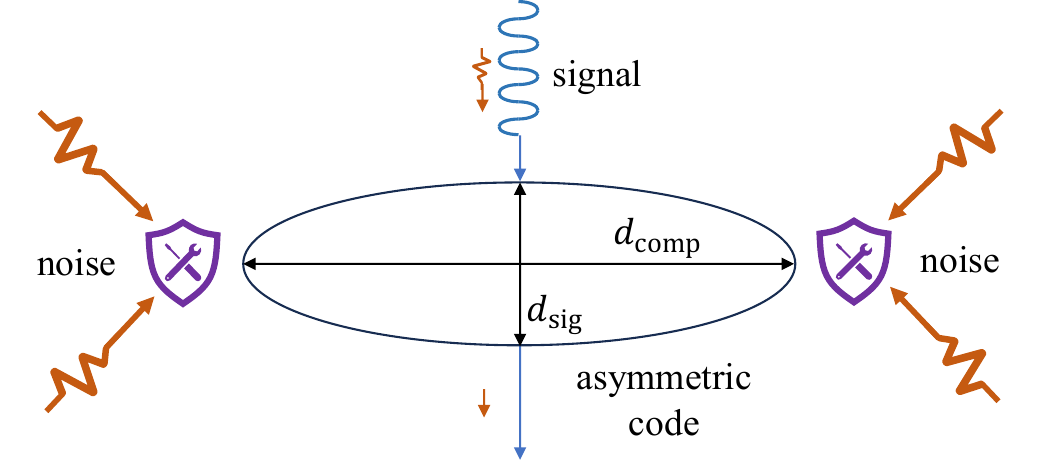}
\caption{}
\label{fig:asymmetriccode}
\end{subfigure}
\hspace{0.01\textwidth}
\begin{subfigure}[t]{0.48\textwidth}
\centering
\includegraphics[width=\linewidth]{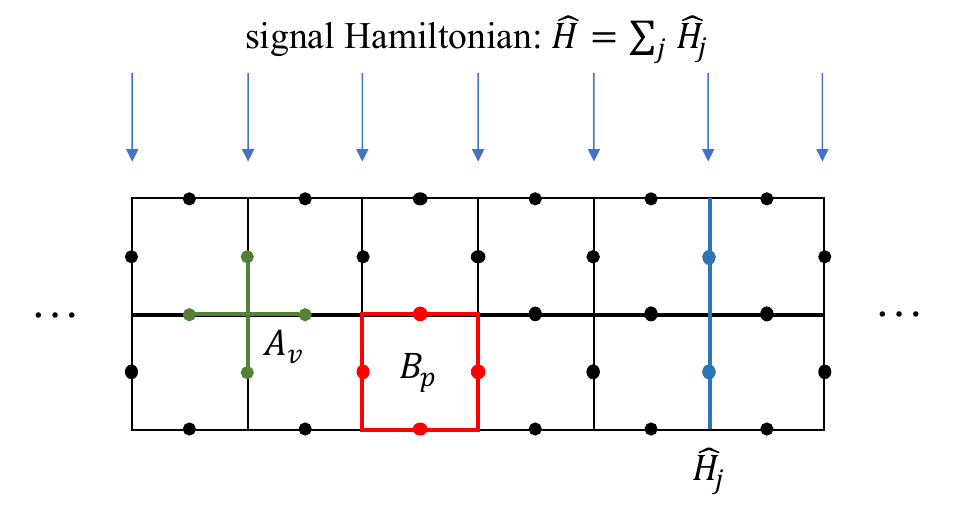}
\caption{}
\label{fig:asymmetrictoric}
\end{subfigure}
\caption{\textbf{Asymmetric protection and an example.}
\textbf{(a)} Asymmetric QEC protects different local directions unequally. The signal direction is assigned a small effective distance $d_{\mathrm{sig}}$, allowing local Hamiltonian components to act as accessible representatives of the same logical signal. The complementary directions are assigned a larger effective distance $d_{\mathrm{comp}}$, so that noise in these directions remains correctable.
\textbf{(b)} Asymmetric toric-code example. A rectangular toric code with highly unequal linear dimensions provides a geometric realization of the same principle. Qubits are placed on edges, with $X$-type star stabilizers $A_v$ and $Z$-type plaquette stabilizers $B_p$. The signal Hamiltonian is chosen as a sum of short string operators $\hat{H}_j$ running across the narrow direction. Each $\hat{H}_j$ acts as a low-weight representative of the same signal-aligned logical operator, while products of different $\hat{H}_j$ are stabilizers. As a result, their contributions to the QFI add coherently, allowing the corresponding code state to achieve Heisenberg-limited scaling while retaining protection in the complementary logical directions.}
\label{fig:asymmetric}
\end{figure}

This observation suggests that the relevant notion of protection for metrology should be direction resolved. The signal direction and its complementary logical directions play different roles: local operators aligned with the signal should remain accessible and act coherently as the same logical observable, whereas local operators in complementary directions should remain detectable and correctable with a growing effective distance, as shown in Fig.~\ref{fig:asymmetriccode}. Related forms of asymmetric protection have appeared in specialized sensing codes, such as repetition- or cat-type constructions for particular signal-noise geometries~\cite{Dur2014improved,Kessler2014quantum}. Here, however, our goal is not to add another isolated example. We develop a general construction principle that, for any local Hamiltonian, builds a stabilizer code adapted to the signal, ensuring that the signal direction has a constant effective distance while complementary directions retain growing protection.

Motivated by this principle, we introduce a class of QEC codes with intrinsically asymmetric protection, which we refer to as asymmetric QEC codes. Although this terminology is inspired by earlier coding-theoretic notions of asymmetric quantum codes, where unequal protection is tailored to different physical error types or error rates~\cite{Ioffe2007asymmetric,Sarvepalli2009asymmetric}, our notion is different: the asymmetry is defined relative to the metrological signal and its complementary logical directions. Formally, we consider stabilizer codes that admit a pair of anti-commuting logical operators $\bar{X}^*$ and $\bar{Z}^*$ with parametrically different effective weights.

\begin{definition}[Asymmetric QEC code]
Consider a family of stabilizer codes on $n$ physical qubits with stabilizer group $\mathbb{S}_n$. We call the family asymmetric if there exists a pair of anti-commuting logical Pauli operators $\bar{X}^*$ and $\bar{Z}^*$, with $\bar{Z}^*$ chosen as the signal-aligned logical direction, such that
\begin{equation}
\mathrm{wt}_{\mathrm{eff}}(\bar{Z}^*)=\Theta(1),\qquad \mathrm{wt}_{\mathrm{eff}}(\bar{X}^*)=\Theta(n).
\end{equation}
Here, the effective weight of a logical operator $\bar{O}$ is defined as $\mathrm{wt}_{\mathrm{eff}}(\bar{O})\equiv\min_{S\in\mathbb{S}_n}\mathrm{wt}(S\bar{O})$.
\end{definition}

This asymmetry captures the basic requirement for protected metrology. The signal-aligned logical operator $\bar{Z}^*$ remains locally accessible, while the conjugate logical direction $\bar{X}^*$ has macroscopic effective weight and is therefore protected against low-weight perturbations. Thus, sensitivity and protection coexist, but along different logical directions.


This direction-resolved design can be implemented systematically and is not tied to a particular sensing geometry. Given a local Hamiltonian, we construct a corresponding asymmetric code adapted to its Pauli decomposition, so that an extensive set of local signal components is made to act as different physical representatives of the same logical operator. The construction, therefore, applies to arbitrary local signals in the setting considered here, rather than providing another isolated repetition- or cat-code example.

\begin{theorem}[General asymmetric construction]\label{theo:construction}
For any $K$-local Hamiltonian defined in Eq.~\eqref{eq:Hdecom}, one can construct an asymmetric quantum error-correcting code $\mathcal{C}_{\hat{H}}$ such that there exists a code state $\ket{\psi}\in\mathcal{C}_{\hat{H}}$ with $\bar{X}^*\ket{\psi}=\ket{\psi}$ achieving Heisenberg-limited scaling of the QFI, $\mathcal{F}(\ket{\psi},\hat{H})=\Omega(n^2)$.
\end{theorem}

The construction starts from the Pauli decomposition of $\hat{H}$, $\hat{H}=\sum_j\alpha_j P_j$, with signs absorbed into $P_j$ so that $\alpha_j\geq 0$. We select an extensive mutually commuting collection of local Pauli terms with total coefficient weight $\Omega(n)$, and impose their pairwise products as stabilizers. Inside the code space, each selected local term is therefore a constant-weight physical representative of the same signal-aligned logical operator $\bar{Z}^*$.

The intuition for both asymmetry and Heisenberg scaling is then direct. Because the selected signal terms all act as $\bar{Z}^*$, their contributions add in phase on a code state polarized along the conjugate direction $\bar{X}^*$, producing an effective logical field of strength $\Omega(n)$ and hence QFI $\Omega(n^2)$. At the same time, any conjugate logical operator must anticommute with an extensive number of these local signal representatives while commuting with their stabilizer products. Bounded-degree locality then forces its effective weight to grow with $n$. Thus, the signal direction remains locally accessible, whereas the complementary logical direction is protected by a macroscopic effective distance. The full construction, including the choice of stabilizer sector and the treatment of Hamiltonian terms outside the selected commuting set, is given in Methods and illustrated in Fig.~\ref{fig:construction}.

Topological codes provide a simple and physically relevant geometric realization of asymmetric QEC, because the weights of their logical string operators can be tuned by the aspect ratio of the lattice. We illustrate this idea with the toric code, although the same principle applies more broadly to surface-code architectures~\cite{Kitaev2003Fault,Fowler2012Surface,Acharya2023Suppressing,Acharya2025Quantum}. Consider a rectangular toric code with periodic boundary conditions, where the short direction has constant length $L_y=c=O(1)$ and the long direction has length $L_x=\frac{n}{2c}=\Theta(n)$, as shown in Fig.~\ref{fig:asymmetrictoric}. Logical strings wrapping around the short cycle then have constant weight, whereas their conjugate strings wrapping around the long cycle have weight proportional to $n$. Choosing a short logical $Z$ string as the signal-aligned operator $\bar{Z}^*$ gives $\mathrm{wt}(\bar{Z}^*)=L_y=O(1)$, while the conjugate logical operator $\bar{X}^*$ has $\mathrm{wt}(\bar{X}^*)=L_x=\Theta(n)$. The code is therefore asymmetric: the signal direction is locally accessible, but the complementary direction remains protected by a macroscopic logical string.

To exploit this geometry for metrology, we choose the signal Hamiltonian to be the sum of all parallel $Z$ strings wrapping around the short cycle. Each term acts on only $c$ physical qubits and is therefore local, while all such strings are stabilizer-equivalent representatives of the same logical operator $\bar{Z}^*$. On a code state polarized in the conjugate basis, $\bar{X}^*\ket{\psi}=\ket{\psi}$, each short string has zero expectation value, but different strings are perfectly correlated. Their contributions therefore add coherently, giving Heisenberg-limited QFI, $\mathcal{F}(\ket{\psi},\hat{H})=\frac{n^2}{4c^2}=\Theta(n^2)$.

This example makes the design principle transparent. Geometric asymmetry leaves one logical direction deliberately low weight so that a local signal can accumulate coherently, while the conjugate direction retains a growing effective distance and hence protection against local perturbations. The toric-code construction is thus a concrete local instance of the general Hamiltonian-adapted asymmetric construction above.

For implementation, the Hamiltonian-adapted construction can be strengthened without changing its metrological mechanism. First, the asymmetry can be made stronger: the signal-aligned logical operator $\bar{Z}^*$ is the only logical operator with a constant-weight representative, while all logical directions independent of $\bar{Z}^*$ have growing effective weight. Thus, the code singles out only the metrological direction, while retaining scalable protection in all complementary directions. Second, it can be made QLDPC-like, with stabilizer generators of bounded weight and each physical qubit involved in only a bounded number of checks. Under the same assumptions as Theorem~\ref{theo:construction}, such strongly asymmetric QLDPC codes can be constructed while preserving the Heisenberg-limited QFI scaling. The completion procedure is given in Methods and Supplementary Information.

Concatenated asymmetric codes further provide a tunable interpolation between protecting the signal direction and retaining metrological gain. For a parameter $L$ with $\sqrt{n}\leq L\le n$, the signal-aligned logical direction has effective distance $\Theta(n/L)$, while complementary logical directions can have effective distance $\Theta(L)$. The corresponding probe states achieve $\mathcal{F}=\Theta(nL)$. Thus, the choice $L=n$ recovers the fully asymmetric regime with Heisenberg-limited scaling and a constant signal-direction distance, whereas smaller $L$ increases protection along the signal direction at the cost of reduced QFI. This family connects our asymmetric construction to generalized Shor-type codes, making the protection-sensitivity trade-off continuously tunable. Details of the concatenated construction and corresponding proof are provided in Methods and Supplementary Information. An overall illustration of properties of asymmetric codes is shown in Fig.~\ref{fig:asymmetricproperty}.

The metrological probe states of the asymmetric codes are also efficiently preparable. For the Hamiltonian-adapted code $\mathcal{C}_{\hat{H}}$, one can prepare a code state $\ket{\psi}\in\mathcal{C}_{\hat{H}}$ satisfying $\bar{X}^*\ket{\psi}=\ket{\psi}$ and $\mathcal{F}(\ket{\psi},\hat{H})=\Omega(n^2)$ using a constant-depth adaptive Clifford circuit with $O(n)$ ancillary qubits. Here, adaptivity refers to the use of intermediate measurements together with feed-forward operations conditioned on their outcomes. Physically, these measurements project the system into the desired stabilizer sector, while feed-forward fixes the stabilizer and logical signs needed for the selected local signal terms to add coherently.

This preparation result makes the asymmetric-code probes not only valid in principle, but scalable in practice: the circuit depth remains independent of system size, and the ancilla overhead is only linear in $n$. We further show in Supplementary Material that this resource scaling is essentially optimal for achieving Heisenberg-limited metrological response in this setting. The concatenated asymmetric code states can likewise be prepared by shallow adaptive circuits, with the corresponding resource estimates given in Methods and Supplementary Information. Consequently, the proposed scheme gives a practical resource blueprint for scalable quantum-enhanced metrology.

\section{Discussion}

In this work, we have shown that QEC-assisted metrology is constrained by a structural protection-sensitivity trade-off. The syndrome structure that protects encoded information from local perturbations can also suppress the coherent accumulation of local signal terms required for quantum-enhanced precision. This incompatibility limits broad classes of symmetrically protected codes to standard-quantum-limit scaling, while degenerate constructions such as generalized Shor codes can only interpolate within the same trade-off. The incompatibility can be bypassed by making the error protection directional. Asymmetric QEC leaves the signal direction locally accessible while protecting complementary directions, thereby restoring Heisenberg-limited sensitivity for arbitrary local sensing Hamiltonians. Together with strongly asymmetric QLDPC constructions, concatenated tunable variants and constant-depth adaptive preparation, these results provide a scalable framework for designing quantum error-correcting code states specifically for metrology, which could in principle be probed using recent QFI-estimation techniques~\cite{Vitale2024Robust}.

A central open question is whether the protection-sensitivity trade-off is universal after additional requirements are included to exclude trivial counterexamples. Our results cover three representative mechanisms by which exact QEC codes protect against local errors, which suggests that the trade-off is not an artifact of a particular code construction. We conjecture that any nontrivial QEC-assisted metrological scheme obeys a bound of the form $d_{\mathrm{sig}}\mathcal{F}=O(n^2)$, or an appropriate variant of it, where $d_{\mathrm{sig}}$ quantifies protection against local perturbations along the same physical directions that carry the signal. For symmetrically protected codes, $d_{\mathrm{sig}}$ reduces to the ordinary code distance; for asymmetric codes, it remains small. The nontriviality requirement is essential: one must exclude constructions in which an unprotected subsystem generates the metrological sensitivity while an unrelated encoded subsystem supplies a large distance. A meaningful formulation should require that the degrees of freedom contributing extensively to the QFI are themselves protected, and should specify the relevant notions of locality, bounded signal strength, and exact, approximate, or biased correctability. Proving such a theorem, or finding a genuine counterexample, would clarify the ultimate compatibility between local-error protection and coherent signal amplification.

Another important direction is to integrate asymmetric QEC into fully fault-tolerant metrological protocols. Our results characterize encoded probe states and their metrological response, but a complete sensing architecture must also protect state preparation, syndrome extraction, recovery, signal evolution, decoding, and feedback against realistic faults. This is particularly subtle for asymmetric codes, because the signal-aligned direction is intentionally left weakly protected: this enables coherent signal accumulation, but also limits which noise components can be corrected without erasing the signal. Developing threshold analyses, finite-rate recovery schemes, and noise-adapted protocols for asymmetric QEC would therefore be an important step toward practical large-scale quantum sensing~\cite{Mann2025Quantum,Liu2025Heisenberg,Kwon2026virtual,sahu2026achieving,Yamamoto2022Error}.

Finally, the measurement stage should be treated on the same importance as state preparation and sensing. Our analysis uses the QFI, which assumes access to an optimal measurement, whereas realistic devices are constrained by noisy, finite-depth, and often local readout. It remains to design readout and estimation procedures that extract the enhanced signal from asymmetric code states while preserving their protection against complementary errors. Combining asymmetric code design with protected logical measurements, adaptive estimation and platform-specific noise models will be crucial for turning the structural advantage identified here into scalable quantum advantage in realistic metrological devices.

\begin{acknowledgements}
We express our gratitude to Yifei Lu and Boren Gu for the insightful discussions. 
J.~C., R.~L., Z.~D., Y.~Y. and X.~M. acknowledge the support from the National Natural Science Foundation of China Grants No.~12174216,  the Innovation Program for Quantum Science and Technology Grant No.~2021ZD0300804,  No.~2021ZD0300702, the CCF-QuantumCtek Superconducting Quantum Computing Special Cooperation Program (Grant No.~CCF-QC2025005), and the Turing AI Institute of Nanjing.
Y.Z. acknowledges the support from the National Natural Science Foundation of China (NSFC) Grant No.~12575012, the Quantum Science and Technology-National Science and Technology Major Project Grant Nos.~2024ZD0301900 and 2021ZD0302000, the Shanghai QiYuan Innovation Foundation, the Shanghai Municipal Commission of Science and Technology with Grant No.~25511103200, the Shanghai Science and Technology Innovation Action Plan Grant No.~24LZ1400200, the Shanghai Pilot Program for Basic Research-Fudan University 21TQ1400100 (25TQ003), the CCF-Quantum CTek Superconducting Quantum Computing CCF-QC2025006.
\end{acknowledgements}

 \bibliographystyle{apsrev}

\bibliography{./tex/QECMetrology1.bib}

@article{Yu2020LIGO,
author={Yu, Haocun and McCuller, L. and Tse, M. and Kijbunchoo, N. and Barsotti, L. and Mavalvala, N. and Betzwieser, J. and Blair, C. D. and Dwyer, S. E. and Effler, A. and Evans, M. and Fernandez-Galiana, A. and Fritschel, P. and Frolov, V. V. and Matichard, F. and McClelland, D. E. and McRae, T. and Mullavey, A. and Sigg, D. and Slagmolen, B. J. J. and Whittle, C. and Buikema, A. and Chen, Y. and Corbitt, T. R. and Schnabel, R. and Abbott, R. and Adams, C. and Adhikari, R. X. and Ananyeva, A. and Appert, S. and Arai, K. and Areeda, J. S. and Asali, Y. and Aston, S. M. and Austin, C. and Baer, A. M. and Ball, M. and Ballmer, S. W. and Banagiri, S. and Barker, D. and Bartlett, J. and Berger, B. K. and Bhattacharjee, D. and Billingsley, G. and Biscans, S. and Blair, R. M. and Bode, N. and Booker, P. and Bork, R. and Bramley, A. and Brooks, A. F. and Brown, D. D. and Cahillane, C. and Cannon, K. C. and Chen, X. and Ciobanu, A. A. and Clara, F. and Cooper, S. J. and Corley, K. R. and Countryman, S. T. and Covas, P. B. and Coyne, D. C. and Datrier, L. E. H. and Davis, D. and Di Fronzo, C. and Dooley, K. L. and Driggers, J. C. and Dupej, P. and Etzel, T. and Evans, T. M. and Feicht, J. and Fulda, P. and Fyffe, M. and Giaime, J. A. and Giardina, K. D. and Godwin, P. and Goetz, E. and Gras, S. and Gray, C. and Gray, R. and Green, A. C. and Gupta, Anchal and Gustafson, E. K. and Gustafson, R. and Hanks, J. and Hanson, J. and Hardwick, T. and Hasskew, R. K. and Heintze, M. C. and Helmling-Cornell, A. F. and Holland, N. A. and Jones, J. D. and Kandhasamy, S. and Karki, S. and Kasprzack, M. and Kawabe, K. and King, P. J. and Kissel, J. S. and Kumar, Rahul and Landry, M. and Lane, B. B. and Lantz, B. and Laxen, M. and Lecoeuche, Y. K. and Leviton, J. and Liu, J. and Lormand, M. and Lundgren, A. P. and Macas, R. and MacInnis, M. and Macleod, D. M.and Mansell, G. L. and M{\'a}rka, S. and M{\'a}rka, Z. and Martynov, D. V. and Mason, K. and Massinger, T. J. and McCarthy, R. and McCormick, S. and McIver, J. and Mendell, G. and Merfeld, K. and Merilh, E. L. and Meylahn, F. and Mistry, T. and Mittleman, R. and Moreno, G. and Mow-Lowry, C. M. and Mozzon, S. and Nelson, T. J. N. and Nguyen, P. and Nuttall, L. K. and Oberling, J. and Oram, Richard J. and Osthelder, C. and Ottaway, D. J. and Overmier, H. and Palamos, J. R. and Parker, W. and Payne, E. and Pele, A. and Perez, C. J. and Pirello, M. and Radkins, H. and Ramirez, K. E. and Richardson, J. W. and Riles, K. and Robertson, N. A. and Rollins, J. G. and Romel, C. L. and Romie, J. H. and Ross, M. P. and Ryan, K. and Sadecki, T. and Sanchez, E. J. and Sanchez, L. E. and Saravanan, T. R. and Savage, R. L. and Schaetzl, D. and Schofield, R. M. S. and Schwartz, E. and Sellers, D. and Shaffer, T. and Smith, J. R. and Soni, S. and Sorazu, B. and Spencer, A. P. and Strain, K. A. and Sun, L. and Szczepa{\'{n}}czyk, M. J. and Thomas, M. and Thomas, P. and Thorne, K. A. and Toland, K. and Torrie, C. I. and Traylor, G. and Urban, A. L. and Vajente, G. and Valdes, G. and Vander-Hyde, D. C. and Veitch, P. J. and Venkateswara, K. and Venugopalan, G. and Viets, A. D. and Vo, T. and Vorvick, C. and Wade, M. and Ward, R. L. and Warner, J. and Weaver, B. and Weiss, R. and Willke, B. and Wipf, C. C. and Xiao, L. and Yamamoto, H. and Yu, Hang and Zhang, L. and Zucker, M. E. and Zweizig, J. and members of the LIGO Scientific Collaboration},
title={Quantum correlations between light and the kilogram-mass mirrors of LIGO},
journal={Nature},
year={2020},
month={Jul},
day={01},
volume={583},
number={7814},
pages={43-47},
issn={1476-4687},
doi={10.1038/s41586-020-2420-8},
url={https://doi.org/10.1038/s41586-020-2420-8}
}

@article{McCormick2019Quantum,
author={McCormick, Katherine C. and Keller, Jonas and Burd, Shaun C. and Wineland, David J. and Wilson, Andrew C. and Leibfried, Dietrich},
title={Quantum-enhanced sensing of a single-ion mechanical oscillator},
journal={Nature},
year={2019},
month={Aug},
day={01},
volume={572},
number={7767},
pages={86-90},
issn={1476-4687},
doi={10.1038/s41586-019-1421-y},
url={https://doi.org/10.1038/s41586-019-1421-y}
}

@article{Franke2023Quantum,
author={Franke, Johannes and Muleady, Sean R. and Kaubruegger, Raphael and Kranzl, Florian and Blatt, Rainer and Rey, Ana Maria and Joshi, Manoj K. and Roos, Christian F.},
title={Quantum-enhanced sensing on optical transitions through finite-range interactions},
journal={Nature},
year={2023},
month={Sep},
day={01},
volume={621},
number={7980},
pages={740-745},
issn={1476-4687},
doi={10.1038/s41586-023-06472-z},
url={https://doi.org/10.1038/s41586-023-06472-z}
}

@article{Giovannetti2004Quantum,
author = {Vittorio Giovannetti and Seth Lloyd and Lorenzo Maccone},
title = {Quantum-Enhanced Measurements: Beating the Standard Quantum Limit},
journal = {Science},
volume = {306},
number = {5700},
pages = {1330-1336},
year = {2004},
doi = {10.1126/science.1104149},
URL = {https://www.science.org/doi/abs/10.1126/science.1104149},
eprint = {https://www.science.org/doi/pdf/10.1126/science.1104149}
}

@article{Giovannetti2006metrology,
  title = {Quantum Metrology},
  author = {Giovannetti, Vittorio and Lloyd, Seth and Maccone, Lorenzo},
  journal = {Phys. Rev. Lett.},
  volume = {96},
  issue = {1},
  pages = {010401},
  numpages = {4},
  year = {2006},
  month = {Jan},
  publisher = {American Physical Society},
  doi = {10.1103/PhysRevLett.96.010401},
  url = {https://link.aps.org/doi/10.1103/PhysRevLett.96.010401}
}

@article{Huelga1997Improvement,
  title = {Improvement of Frequency Standards with Quantum Entanglement},
  author = {Huelga, S. F. and Macchiavello, C. and Pellizzari, T. and Ekert, A. K. and Plenio, M. B. and Cirac, J. I.},
  journal = {Phys. Rev. Lett.},
  volume = {79},
  issue = {20},
  pages = {3865--3868},
  numpages = {0},
  year = {1997},
  month = {Nov},
  publisher = {American Physical Society},
  doi = {10.1103/PhysRevLett.79.3865},
  url = {https://link.aps.org/doi/10.1103/PhysRevLett.79.3865}
}

@article{Escher2011General,
author={Escher, B. M. and de Matos Filho, R. L. and Davidovich, L.},
title={General framework for estimating the ultimate precision limit in noisy quantum-enhanced metrology},
journal={Nature Physics},
year={2011},
month={May},
day={01},
volume={7},
number={5},
pages={406-411},
issn={1745-2481},
doi={10.1038/nphys1958},
url={https://doi.org/10.1038/nphys1958}
}

@article{Demkowicz2012elusive,
author={Demkowicz-Dobrza{\'{n}}ski, Rafa{\l} and Ko{\l}ody{\'{n}}ski, Jan and Gu{\c{T}}{\u{a}}, M{\u{a}}d{\u{a}}lin},
title={The elusive Heisenberg limit in quantum-enhanced metrology},
journal={Nature Communications},
year={2012},
month={Sep},
day={18},
volume={3},
number={1},
pages={1063},
issn={2041-1723},
doi={10.1038/ncomms2067},
url={https://doi.org/10.1038/ncomms2067}
}

@article{Kessler2014quantum,
  title = {Quantum Error Correction for Metrology},
  author = {Kessler, E. M. and Lovchinsky, I. and Sushkov, A. O. and Lukin, M. D.},
  journal = {Phys. Rev. Lett.},
  volume = {112},
  issue = {15},
  pages = {150802},
  numpages = {5},
  year = {2014},
  month = {Apr},
  publisher = {American Physical Society},
  doi = {10.1103/PhysRevLett.112.150802},
  url = {https://link.aps.org/doi/10.1103/PhysRevLett.112.150802}
}

@article{Dur2014improved,
  title = {Improved Quantum Metrology Using Quantum Error Correction},
  author = {D\"ur, W. and Skotiniotis, M. and Fr\"owis, F. and Kraus, B.},
  journal = {Phys. Rev. Lett.},
  volume = {112},
  issue = {8},
  pages = {080801},
  numpages = {5},
  year = {2014},
  month = {Feb},
  publisher = {American Physical Society},
  doi = {10.1103/PhysRevLett.112.080801},
  url = {https://link.aps.org/doi/10.1103/PhysRevLett.112.080801}
}

@article{zhou2018achieving,
author={Zhou, Sisi and Zhang, Mengzhen and Preskill, John and Jiang, Liang},
title={Achieving the Heisenberg limit in quantum metrology using quantum error correction},
journal={Nature Communications},
year={2018},
month={Jan},
day={08},
volume={9},
number={1},
pages={78},
issn={2041-1723},
doi={10.1038/s41467-017-02510-3},
url={https://doi.org/10.1038/s41467-017-02510-3}
}

@article{Stray2022gravity,
author={Stray, Ben and Lamb, Andrew and Kaushik, Aisha and Vovrosh, Jamie and Rodgers, Anthony and Winch, Jonathan and Hayati, Farzad and Boddice, Daniel and Stabrawa, Artur and Niggebaum, Alexander and Langlois, Mehdi and Lien, Yu-Hung and Lellouch, Samuel and Roshanmanesh, Sanaz and Ridley, Kevin and de Villiers, Geoffrey and Brown, Gareth and Cross, Trevor and Tuckwell, George and Faramarzi, Asaad and Metje, Nicole and Bongs, Kai and Holynski, Michael},
title={Quantum sensing for gravity cartography},
journal={Nature},
year={2022},
month={Feb},
day={01},
volume={602},
number={7898},
pages={590-594},
issn={1476-4687},
doi={10.1038/s41586-021-04315-3},
url={https://doi.org/10.1038/s41586-021-04315-3}
}

@article{Roslund2024clock,
author={Roslund, Jonathan D. and Cing{\"o}z, Arman and Lunden, William D. and Partridge, Guthrie B. and Kowligy, Abijith S. and Roller, Frank and Sheredy, Daniel B. and Skulason, Gunnar E. and Song, Joe P. and Abo-Shaeer, Jamil R. and Boyd, Martin M.},
title={Optical clocks at sea},
journal={Nature},
year={2024},
month={Apr},
day={01},
volume={628},
number={8009},
pages={736-740},
issn={1476-4687},
doi={10.1038/s41586-024-07225-2},
url={https://doi.org/10.1038/s41586-024-07225-2}
}

@article{Malia2022Distributed,
author={Malia, Benjamin K. and Wu, Yunfan and Mart{\'i}nez-Rinc{\'o}n, Juli{\'a}n and Kasevich, Mark A.},
title={Distributed quantum sensing with mode-entangled spin-squeezed atomic states},
journal={Nature},
year={2022},
month={Dec},
day={01},
volume={612},
number={7941},
pages={661-665},
issn={1476-4687},
doi={10.1038/s41586-022-05363-z},
url={https://doi.org/10.1038/s41586-022-05363-z}
}

@article{Terhel2015QEC,
  title = {Quantum error correction for quantum memories},
  author = {Terhal, Barbara M.},
  journal = {Rev. Mod. Phys.},
  volume = {87},
  issue = {2},
  pages = {307--346},
  numpages = {40},
  year = {2015},
  month = {Apr},
  publisher = {American Physical Society},
  doi = {10.1103/RevModPhys.87.307},
  url = {https://link.aps.org/doi/10.1103/RevModPhys.87.307}
}

@article{Demkowicz2017Adaptive,
  title = {Adaptive Quantum Metrology under General Markovian Noise},
  author = {Demkowicz-Dobrza\ifmmode \acute{n}\else \'{n}\fi{}ski, Rafa\l{} and Czajkowski, Jan and Sekatski, Pavel},
  journal = {Phys. Rev. X},
  volume = {7},
  issue = {4},
  pages = {041009},
  numpages = {15},
  year = {2017},
  month = {Oct},
  publisher = {American Physical Society},
  doi = {10.1103/PhysRevX.7.041009},
  url = {https://link.aps.org/doi/10.1103/PhysRevX.7.041009}
}

@article{Layden2019Ancilla,
  title = {Ancilla-Free Quantum Error Correction Codes for Quantum Metrology},
  author = {Layden, David and Zhou, Sisi and Cappellaro, Paola and Jiang, Liang},
  journal = {Phys. Rev. Lett.},
  volume = {122},
  issue = {4},
  pages = {040502},
  numpages = {6},
  year = {2019},
  month = {Jan},
  publisher = {American Physical Society},
  doi = {10.1103/PhysRevLett.122.040502},
  url = {https://link.aps.org/doi/10.1103/PhysRevLett.122.040502}
}

@article{Rojkov2022Bias,
  title = {Bias in Error-Corrected Quantum Sensing},
  author = {Rojkov, Ivan and Layden, David and Cappellaro, Paola and Home, Jonathan and Reiter, Florentin},
  journal = {Phys. Rev. Lett.},
  volume = {128},
  issue = {14},
  pages = {140503},
  numpages = {6},
  year = {2022},
  month = {Apr},
  publisher = {American Physical Society},
  doi = {10.1103/PhysRevLett.128.140503},
  url = {https://link.aps.org/doi/10.1103/PhysRevLett.128.140503}
}

@article{Mann2025Quantum,
  title = {Quantum Error-Corrected Non-Markovian Metrology},
  author = {Mann, Zachary and Cao, Ningping and Laflamme, Raymond and Zhou, Sisi},
  journal = {PRX Quantum},
  volume = {6},
  issue = {3},
  pages = {030321},
  numpages = {31},
  year = {2025},
  month = {Aug},
  publisher = {American Physical Society},
  doi = {10.1103/wfyl-wtz3},
  url = {https://link.aps.org/doi/10.1103/wfyl-wtz3}
}

@article{Braunstein1994statistical,
  title = {Statistical distance and the geometry of quantum states},
  author = {Braunstein, Samuel L. and Caves, Carlton M.},
  journal = {Phys. Rev. Lett.},
  volume = {72},
  issue = {22},
  pages = {3439--3443},
  numpages = {0},
  year = {1994},
  month = {May},
  publisher = {American Physical Society},
  doi = {10.1103/PhysRevLett.72.3439},
  url = {https://link.aps.org/doi/10.1103/PhysRevLett.72.3439}
}

@article{Paris2009quantum,
author = {Paris, Matteo G. A.},
title = {QUANTUM ESTIMATION FOR QUANTUM TECHNOLOGY},
journal = {International Journal of Quantum Information},
volume = {07},
number = {supp01},
pages = {125-137},
year = {2009},
doi = {10.1142/S0219749909004839},
URL = {https://doi.org/10.1142/S0219749909004839}
}

@article{Knill1997KLcondition,
  title = {Theory of quantum error-correcting codes},
  author = {Knill, Emanuel and Laflamme, Raymond},
  journal = {Phys. Rev. A},
  volume = {55},
  issue = {2},
  pages = {900--911},
  numpages = {0},
  year = {1997},
  month = {Feb},
  publisher = {American Physical Society},
  doi = {10.1103/PhysRevA.55.900},
  url = {https://link.aps.org/doi/10.1103/PhysRevA.55.900}
}

@misc{Gottesman2014Faulttolerant,
      title={Fault-Tolerant Quantum Computation with Constant Overhead}, 
      author={Daniel Gottesman},
      year={2014},
      eprint={1310.2984},
      archivePrefix={arXiv}
}

@inproceedings{Dinur2023GoodQLDPC,
author = {Dinur, Irit and Hsieh, Min-Hsiu and Lin, Ting-Chun and Vidick, Thomas},
title = {Good Quantum LDPC Codes with Linear Time Decoders},
year = {2023},
isbn = {9781450399135},
publisher = {Association for Computing Machinery},
address = {New York, NY, USA},
url = {https://doi.org/10.1145/3564246.3585101},
doi = {10.1145/3564246.3585101},
booktitle = {Proceedings of the 55th Annual ACM Symposium on Theory of Computing},
pages = {905–918},
numpages = {14},
location = {Orlando, FL, USA},
series = {STOC 2023}
}

@article{Bravyi2024FTmemory,
author={Bravyi, Sergey and Cross, Andrew W. and Gambetta, Jay M. and Maslov, Dmitri and Rall, Patrick and Yoder, Theodore J.},
title={High-threshold and low-overhead fault-tolerant quantum memory},
journal={Nature},
year={2024},
month={Mar},
day={01},
volume={627},
number={8005},
pages={778-782},
issn={1476-4687},
doi={10.1038/s41586-024-07107-7},
url={https://doi.org/10.1038/s41586-024-07107-7}
}

@article{Ioffe2007asymmetric,
  title = {Asymmetric quantum error-correcting codes},
  author = {Ioffe, Lev and M\'ezard, Marc},
  journal = {Phys. Rev. A},
  volume = {75},
  issue = {3},
  pages = {032345},
  numpages = {4},
  year = {2007},
  month = {Mar},
  publisher = {American Physical Society},
  doi = {10.1103/PhysRevA.75.032345},
  url = {https://link.aps.org/doi/10.1103/PhysRevA.75.032345}
}

@article{Sarvepalli2009asymmetric,
    author = {Sarvepalli, Pradeep Kiran and Klappenecker, Andreas and Rötteler, Martin},
    title = {Asymmetric quantum codes: constructions, bounds and performance},
    journal = {Proceedings of the Royal Society A: Mathematical, Physical and Engineering Sciences},
    volume = {465},
    number = {2105},
    pages = {1645-1672},
    year = {2009},
    month = {03},
    issn = {1364-5021},
    doi = {10.1098/rspa.2008.0439},
    url = {https://doi.org/10.1098/rspa.2008.0439},
    eprint = {https://royalsocietypublishing.org/rspa/article-pdf/465/2105/1645/755212/rspa.2008.0439.pdf},
}

@article{Kitaev2003Fault,
title = "Fault-tolerant quantum computation by anyons",
journal = "Annals of Physics",
volume = "303",
number = "1",
pages = "2 - 30",
year = "2003",
issn = "0003-4916",
doi = "https://doi.org/10.1016/S0003-4916(02)00018-0",
url = "http://www.sciencedirect.com/science/article/pii/S0003491602000180",
author = "A.Yu. Kitaev"
}

@article{Fowler2012Surface,
  title = {Surface codes: Towards practical large-scale quantum computation},
  author = {Fowler, Austin G. and Mariantoni, Matteo and Martinis, John M. and Cleland, Andrew N.},
  journal = {Phys. Rev. A},
  volume = {86},
  issue = {3},
  pages = {032324},
  numpages = {48},
  year = {2012},
  month = {Sep},
  publisher = {American Physical Society},
  doi = {10.1103/PhysRevA.86.032324},
  url = {https://link.aps.org/doi/10.1103/PhysRevA.86.032324}
}

@article{Acharya2023Suppressing,
author={Acharya, Rajeev and Aleiner, Igor and Allen, Richard and Andersen, Trond I. and Ansmann, Markus and Arute, Frank and Arya, Kunal and Asfaw, Abraham and Atalaya, Juan and Babbush, Ryan and Bacon, Dave and Bardin, Joseph C. and Basso, Joao and Bengtsson, Andreas and Boixo, Sergio and Bortoli, Gina and Bourassa, Alexandre and Bovaird, Jenna and Brill, Leon and Broughton, Michael and Buckley, Bob B. and Buell, David A. and Burger, Tim and Burkett, Brian and Bushnell, Nicholas and Chen, Yu and Chen, Zijun and Chiaro, Ben and Cogan, Josh and Collins, Roberto and Conner, Paul and Courtney, William and Crook, Alexander L. and Curtin, Ben and Debroy, Dripto M. and Del Toro Barba, Alexander and Demura, Sean and Dunsworth, Andrew and Eppens, Daniel and Erickson, Catherine and Faoro, Lara and Farhi, Edward and Fatemi, Reza and Flores Burgos, Leslie and Forati, Ebrahim and Fowler, Austin G. and Foxen, Brooks and Giang, William and Gidney, Craig and Gilboa, Dar and Giustina, Marissa and Grajales Dau, Alejandro and Gross, Jonathan A. and Habegger, Steve and Hamilton, Michael C. and Harrigan, Matthew P. and Harrington, Sean D. and Higgott, Oscar and Hilton, Jeremy and Hoffmann, Markus and Hong, Sabrina and Huang, Trent and Huff, Ashley and Huggins, William J. and Ioffe, Lev B. and Isakov, Sergei V. and Iveland, Justin and Jeffrey, Evan and Jiang, Zhang and Jones, Cody and Juhas, Pavol and Kafri, Dvir and Kechedzhi, Kostyantyn and Kelly, Julian and Khattar, Tanuj and Khezri, Mostafa and Kieferov{\'a}, M{\'a}ria and Kim, Seon and Kitaev, Alexei and Klimov, Paul V. and Klots, Andrey R. and Korotkov, Alexander N. and Kostritsa, Fedor and Kreikebaum, John Mark and Landhuis, David and Laptev, Pavel and Lau, Kim-Ming and Laws, Lily and Lee, Joonho and Lee, Kenny and Lester, Brian J. and Lill, Alexander and Liu, Wayne and Locharla, Aditya and Lucero, Erik and Malone, Fionn D. and Marshall, Jeffrey and Martin, Orio and McClean, Jarrod R and McCourt, Trevo and McEwen, Mat and Megrant, Anthon and Meurer Costa, Bernard and Mi, Xiao and Miao, Kevin C. and Mohseni, Masoud and Montazeri, Shirin and Morvan, Alexis and Mount, Emily and Mruczkiewicz, Wojciech and Naaman, Ofer and Neeley, Matthew and Neill, Charles and Nersisyan, Ani and Neven, Hartmut and Newman, Michael and Ng, Jiun How and Nguyen, Anthony and Nguyen, Murray and Niu, Murphy Yuezhen and O'Brien, Thomas E. and Opremcak, Alex and Platt, John and Petukhov, Andre and Potter, Rebecca and Pryadko, Leonid P. and Quintana, Chris and Roushan, Pedram and Rubin, Nicholas C. and Saei, Negar and Sank, Daniel and Sankaragomathi, Kannan and Satzinger, Kevin J. and Schurkus, Henry F. and Schuster, Christopher and Shearn, Michael J. and Shorter, Aaron and Shvarts, Vladimir and Skruzny, Jindra and Smelyanskiy, Vadim and Smith, W. Clarke and Sterling, George and Strain, Doug and Szalay, Marco and Torres, Alfredo and Vidal, Guifre and Villalonga, Benjamin and Vollgraff Heidweiller, Catherine and White, Theodore and Xing, Cheng and Yao, Z. Jamie and Yeh, Ping and Yoo, Juhwan and Young, Grayson and Zalcman, Adam and Zhang, Yaxing and Zhu, Ningfeng and AI, Google Quantum},
title={Suppressing quantum errors by scaling a surface code logical qubit},
journal={Nature},
year={2023},
month={Feb},
day={01},
volume={614},
number={7949},
pages={676-681},
issn={1476-4687},
doi={10.1038/s41586-022-05434-1},
url={https://doi.org/10.1038/s41586-022-05434-1}
}

@article{Acharya2025Quantum,
author={Acharya, Rajeev and Abanin, Dmitry A. and Aghababaie-Beni, Laleh and Aleiner, Igor and Andersen, Trond I. and Ansmann, Markus and Arute, Frank and Arya, Kunal and Asfaw, Abraham and Astrakhantsev, Nikita and Atalaya, Juan and Babbush, Ryan and Bacon, Dave and Ballard, Brian and Bardin, Joseph C. and Bausch, Johannes and Bengtsson, Andreas and Bilmes, Alexander and Blackwell, Sam and Boixo, Sergio and Bortoli, Gina and Bourassa, Alexandre and Bovaird, Jenna and Brill, Leon and Broughton, Michael and Browne, David A. and Buchea, Brett and Buckley, Bob B. and Buell, David A. and Burger, Tim and Burkett, Brian and Bushnell, Nicholas and Cabrera, Anthony and Campero, Juan and Chang, Hung-Shen and Chen, Yu and Chen, Zijun and Chiaro, Ben and Chik, Desmond and Chou, Charina and Claes, Jahan and Cleland, Agnetta Y. and Cogan, Josh and Collins, Roberto and Conner, Paul and Courtney, William and Crook, Alexander L. and Curtin, Ben and Das, Sayan and Davies, Alex and De Lorenzo, Laura and Debroy, Dripto M. and Demura, Sean and Devoret, Michel and Di Paolo, Agustin and Donohoe, Paul and Drozdov, Ilya and Dunsworth, Andrew and Earle, Clint and Edlich, Thomas and Eickbusch, Alec and Elbag, Aviv Moshe and Elzouka, Mahmoud and Erickson, Catherine and Faoro, Lara and Farhi, Edward and Ferreira, Vinicius S. and Burgos, Leslie Flore and Forati, Ebrahi and Fowler, Austin G and Foxen, Brook and Ganjam, Suha and Garcia, Gonzal and Gasca, Rober and Genois, {\'E}li and Giang, Willia and Gidney, Crai and Gilboa, Da and Gosula, Raj and Dau, Alejandro Grajale and Graumann, Dietric and Greene, Ale and Gross, Jonathan A and Habegger, Stev and Hall, Joh and Hamilton, Michael C and Hansen, Monic and Harrigan, Matthew P and Harrington, Sean D and Heras, Francisco J. H and Heslin, Stephe and Heu, Paul and Higgott, Osca and Hill, Gordo and Hilton, Jerem and Holland, Georg and Hong, Sabrin and Huang, Hsin-Yua and Huff, Ashle and Huggins, William J and Ioffe, Lev B and Isakov, Sergei V and Iveland, Justi and Jeffrey, Eva and Jiang, Zhan and Jones, Cod and Jordan, Stephe and Joshi, Chaital and Juhas, Pavo and Kafri, Dvi and Kang, Hu and Karamlou, Amir H and Kechedzhi, Kostyanty and Kelly, Julia and Khaire, Trupt and Khattar, Tanuj and Khezri, Mostafa and Kim, Seon and Klimov, Paul V. and Klots, Andrey R. and Kobrin, Bryce and Kohli, Pushmeet and Korotkov, Alexander N. and Kostritsa, Fedor and Kothari, Robin and Kozlovskii, Borislav and Kreikebaum, John Mark and Kurilovich, Vladislav D. and Lacroix, Nathan and Landhuis, David and Lange-Dei, Tiano and Langley, Brandon W. and Laptev, Pavel and Lau, Kim-Ming and Le Guevel, Lo{\"i}ck and Ledford, Justin and Lee, Joonho and Lee, Kenny and Lensky, Yuri D. and Leon, Shannon and Lester, Brian J. and Li, Wing Yan and Li, Yin and Lill, Alexander T. and Liu, Wayne and Livingston, William P. and Locharla, Aditya and Lucero, Erik and Lundahl, Daniel and Lunt, Aaron and Madhuk, Sid and Malone, Fionn D. and Maloney, Ashley and Mandr{\`a}, Salvatore and Manyika, James and Martin, Leigh S. and Martin, Orion and Martin, Steven and Maxfield, Cameron and McClean, Jarrod R. and McEwen, Matt and Meeks, Seneca and Megrant, Anthony and Mi, Xiao and Miao, Kevin C. and Mieszala, Amanda and Molavi, Reza and Molina, Sebastian and Montazeri, Shirin and Morvan, Alexis and Movassagh, Ramis and Mruczkiewicz, Wojciech and Naaman, Ofer and Neeley, Matthew and Neill, Charles and Nersisyan, Ani and Neven, Hartmut and Newman, Michael and Ng, Jiun How and Nguyen, Anthony and Nguyen, Murray and Ni, Chia-Hung and Niu, Murphy Yuezhen and O'Brien, Thomas E. and Oliver, William D. and Opremcak, Alex and Ottosson, Kristoffer and Petukhov, Andre and Pizzuto, Alex and Platt, John and Potter, Rebecca and Pritchard, Orion and Pryadko, Leonid P. and Quintana, Chris and Ramachandran, Ganesh and Reagor, Matthew J. and Redding, John and Rhodes, David M. and Roberts, Gabrielle and Rosenberg, Eliott and Rosenfeld, Emma and Roushan, Pedram and Rubin, Nicholas C and Saei, Nega and Sank, Danie and Sankaragomathi, Kanna and Satzinger, Kevin J and Schurkus, Henry F and Schuster, Christophe and Senior, Andrew W and Shearn, Michael J and Shorter, Aaro and Shutty, Noa and Shvarts, Vladimi and Singh, Shraddh and Sivak, Volodymy and Skruzny, Jindra and Small, Spencer and Smelyanskiy, Vadim and Smith, W. Clarke and Somma, Rolando D. and Springer, Sofia and Sterling, George and Strain, Doug and Suchard, Jordan and Szasz, Aaron and Sztein, Alex and Thor, Douglas and Torres, Alfredo and Torunbalci, M. Mert and Vaishnav, Abeer and Vargas, Justin and Vdovichev, Sergey and Vidal, Guifre and Villalonga, Benjamin and Heidweiller, Catherine Vollgraff and Waltman, Steven and Wang, Shannon X. and Ware, Brayden and Weber, Kate and Weidel, Travis and White, Theodore and Wong, Kristi and Woo, Bryan W. K. and Xing, Cheng and Yao, Z. Jamie and Yeh, Ping and Ying, Bicheng and Yoo, Juhwan and Yosri, Noureldin and Young, Grayson and Zalcman, Adam and Zhang, Yaxing and Zhu, Ningfeng and Zobrist, Nicholas and AI, Google Quantum and {Collaborators}},
title={Quantum error correction below the surface code threshold},
journal={Nature},
year={2025},
month={Feb},
day={01},
volume={638},
number={8052},
pages={920-926},
issn={1476-4687},
doi={10.1038/s41586-024-08449-y},
url={https://doi.org/10.1038/s41586-024-08449-y}
}

@article{Kwon2026virtual,
author={Kwon, Hyukgun and Oh, Changhun and Lim, Youngrong and Jeong, Hyunseok and Lee, Seung-Woo and Jiang, Liang},
title={Virtual purification complements quantum error correction in quantum metrology},
journal={npj Quantum Information},
year={2026},
month={Apr},
day={09},
issn={2056-6387},
doi={10.1038/s41534-026-01231-0},
url={https://doi.org/10.1038/s41534-026-01231-0}
}

@misc{sahu2026achieving,
      title={Achieving the Heisenberg limit using fault-tolerant quantum error correction}, 
      author={Himanshu Sahu and Qian Xu and Sisi Zhou},
      year={2026},
      eprint={2601.05457},
      archivePrefix={arXiv},
      primaryClass={quant-ph},
      url={https://arxiv.org/abs/2601.05457}, 
}

@misc{bjork2009efficiency,
      title={On the efficiency of nondegenerate quantum error correction codes for Pauli channels}, 
      author={Gunnar Bjork and Jonas Almlof and Isabel Sainz},
      year={2009},
      eprint={0810.0541},
      archivePrefix={arXiv},
      primaryClass={quant-ph},
      url={https://arxiv.org/abs/0810.0541}, 
}

@inproceedings{Panteleev2022GoodLDPC,
author = {Panteleev, Pavel and Kalachev, Gleb},
title = {Asymptotically good Quantum and locally testable classical LDPC codes},
year = {2022},
isbn = {9781450392648},
publisher = {Association for Computing Machinery},
address = {New York, NY, USA},
url = {https://doi.org/10.1145/3519935.3520017},
doi = {10.1145/3519935.3520017},
booktitle = {Proceedings of the 54th Annual ACM SIGACT Symposium on Theory of Computing},
pages = {375–388},
numpages = {14},
location = {Rome, Italy},
series = {STOC 2022}
}

@article{Zhang2025Time,
  title = {Time-Efficient Logical Operations on Quantum Low-Density Parity Check Codes},
  author = {Zhang, Guo and Li, Ying},
  journal = {Phys. Rev. Lett.},
  volume = {134},
  issue = {7},
  pages = {070602},
  numpages = {6},
  year = {2025},
  month = {Feb},
  publisher = {American Physical Society},
  doi = {10.1103/PhysRevLett.134.070602},
  url = {https://link.aps.org/doi/10.1103/PhysRevLett.134.070602}
}

@article{Gottesman1999Demonstrating,
author={Gottesman, Daniel
and Chuang, Isaac L.},
title={Demonstrating the viability of universal quantum computation using teleportation and single-qubit operations},
journal={Nature},
year={1999},
month={Nov},
day={01},
volume={402},
number={6760},
pages={390-393},
issn={1476-4687},
doi={10.1038/46503},
url={https://doi.org/10.1038/46503}
}

@article{Raussendorf2002Computational,
author = {Raussendorf, Robert and Briegel, Hans J.},
title = {Computational model underlying the one-way quantum computer},
year = {2002},
issue_date = {October 2002},
publisher = {Rinton Press, Incorporated},
address = {Paramus, NJ},
volume = {2},
number = {6},
issn = {1533-7146},
journal = {Quantum Info. Comput.},
month = oct,
pages = {443–486},
numpages = {44}
}

@article{Bacon2006Operator,
  title = {Operator quantum error-correcting subsystems for self-correcting quantum memories},
  author = {Bacon, Dave},
  journal = {Phys. Rev. A},
  volume = {73},
  issue = {1},
  pages = {012340},
  numpages = {13},
  year = {2006},
  month = {Jan},
  publisher = {American Physical Society},
  doi = {10.1103/PhysRevA.73.012340},
  url = {https://link.aps.org/doi/10.1103/PhysRevA.73.012340}
}

@article{Liu2025Heisenberg,
  title = {Heisenberg-Limited Quantum Metrology without Ancillae},
  author = {Liu, Qiushi and Yang, Yuxiang},
  journal = {Phys. Rev. Lett.},
  volume = {135},
  issue = {14},
  pages = {140801},
  numpages = {8},
  year = {2025},
  month = {Oct},
  publisher = {American Physical Society},
  doi = {10.1103/vmd7-twd5},
  url = {https://link.aps.org/doi/10.1103/vmd7-twd5}
}

@article{Yamamoto2022Error,
  title = {Error-Mitigated Quantum Metrology via Virtual Purification},
  author = {Yamamoto, Kaoru and Endo, Suguru and Hakoshima, Hideaki and Matsuzaki, Yuichiro and Tokunaga, Yuuki},
  journal = {Phys. Rev. Lett.},
  volume = {129},
  issue = {25},
  pages = {250503},
  numpages = {6},
  year = {2022},
  month = {Dec},
  publisher = {American Physical Society},
  doi = {10.1103/PhysRevLett.129.250503},
  url = {https://link.aps.org/doi/10.1103/PhysRevLett.129.250503}
}

@article{Vitale2024Robust,
  title = {Robust Estimation of the Quantum Fisher Information on a Quantum Processor},
  author = {Vitale, Vittorio and Rath, Aniket and Jurcevic, Petar and Elben, Andreas and Branciard, Cyril and Vermersch, Beno\^{\i}t},
  journal = {PRX Quantum},
  volume = {5},
  issue = {3},
  pages = {030338},
  numpages = {27},
  year = {2024},
  month = {Aug},
  publisher = {American Physical Society},
  doi = {10.1103/PRXQuantum.5.030338},
  url = {https://link.aps.org/doi/10.1103/PRXQuantum.5.030338}
}

@misc{liu2025state,
      title={State complexity and phase identification in adaptive quantum circuits}, 
      author={Guoding Liu and Junjie Chen and Xiongfeng Ma},
      year={2025},
      eprint={2509.17014},
      archivePrefix={arXiv},
      primaryClass={quant-ph},
      url={https://arxiv.org/abs/2509.17014}, 
}

\section{Methods}

Here, we present the constructions of asymmetric codes used in the main text, including general asymmetric QEC codes, strongly asymmetric QLDPC codes and concatenated asymmetric codes. It is convenient to begin with a Pauli-basis decomposition of the sensing Hamiltonian. Since each $K$-local term $\hat{H}_j$ acts nontrivially on at most $K$ qubits, we write
\begin{equation}
\hat{H}=\sum_{j=1}^{m}\hat{H}_j=\sum_{a=1}^{M}\alpha_j P_j,
\end{equation}
where each $P_j$ is a Pauli operator of weight at most $K$, $M\leq m\cdot 4^K$, and $|\alpha_j|\leq 1$. By absorbing the signs of the coefficients into the corresponding Pauli operators, we take $\alpha_j\geq 0$ for all $j$ without loss of generality.

For a pure probe state $\ket{\psi}$, the QFI, with the convention used throughout this work, is the variance of the sensing Hamiltonian. It can therefore be decomposed into covariances, or equivalently connected two-point correlation functions, between the Hamiltonian terms. Expanding the Hamiltonian in the Pauli basis gives
\begin{equation}\label{eq:QFIdecomPauli}
\begin{aligned}
\mathcal{F}(\ket{\psi},\hat{H})&=\sum_{j,j'=1}^m\left(\bra{\psi}\hat{H}_j\hat{H}_{j'}\ket{\psi}-\bra{\psi}\hat{H}_j\ket{\psi}\bra{\psi}\hat{H}_{j'}\ket{\psi}\right)\\
&=\sum_{j_1,j_2=1}^M\alpha_{j_1}\alpha_{j_2}\left(\bra{\psi}P_{j_1}P_{j_2}\ket{\psi}-\bra{\psi}P_{j_1}\ket{\psi}\bra{\psi}P_{j_2}\ket{\psi}\right).
\end{aligned}
\end{equation}
This expression makes explicit the mechanism by which stabilizer correlations can enhance metrological sensitivity. For a stabilizer state $\ket{\psi}$, the expectation value of a Pauli operator vanishes unless the operator belongs to the stabilizer group up to a sign, in which case the expectation value is $\pm 1$. Hence, a covariance term in Eq.~\eqref{eq:QFIdecomPauli} can be nonzero only when $P_{j_1}P_{j_2}$ lies in the stabilizer group up to a phase while neither $P_{j_1}$ nor $P_{j_2}$ is itself a stabilizer. In the constructions below, we select pairs for which $P_{j_1}P_{j_2}$ is a Hermitian stabilizer and choose the corresponding stabilizer signs so that these covariance terms contribute positively. This provides the basic design principle for asymmetric codes: local Pauli components of the signal are made indistinguishable within the probe state, so that their contributions add coherently to the QFI, while complementary local directions remain protected by the code.

\subsection{Construction of asymmetric QEC codes}

The construction is guided by the covariance decomposition in Eq.~\eqref{eq:QFIdecomPauli}. A pair of Pauli components $P_{j_1},P_{j_2}$ contributes coherently to the QFI of a stabilizer probe state when $P_{j_1}P_{j_2}$ is a stabilizer while neither $P_{j_1}$ nor $P_{j_2}$ is itself a stabilizer. We therefore aim to choose many local Pauli components of $\hat{H}$ and make them different physical representatives of the same logical operator. Then their pairwise products become stabilizers, whereas the individual operators remain logical and have vanishing expectation value in a conjugate logical eigenstate. This is the basic mechanism behind the construction: the signal-aligned local terms are made indistinguishable within the code space, so their covariance contributions add constructively.

\begin{figure}
\centering
\includegraphics[width=\linewidth]{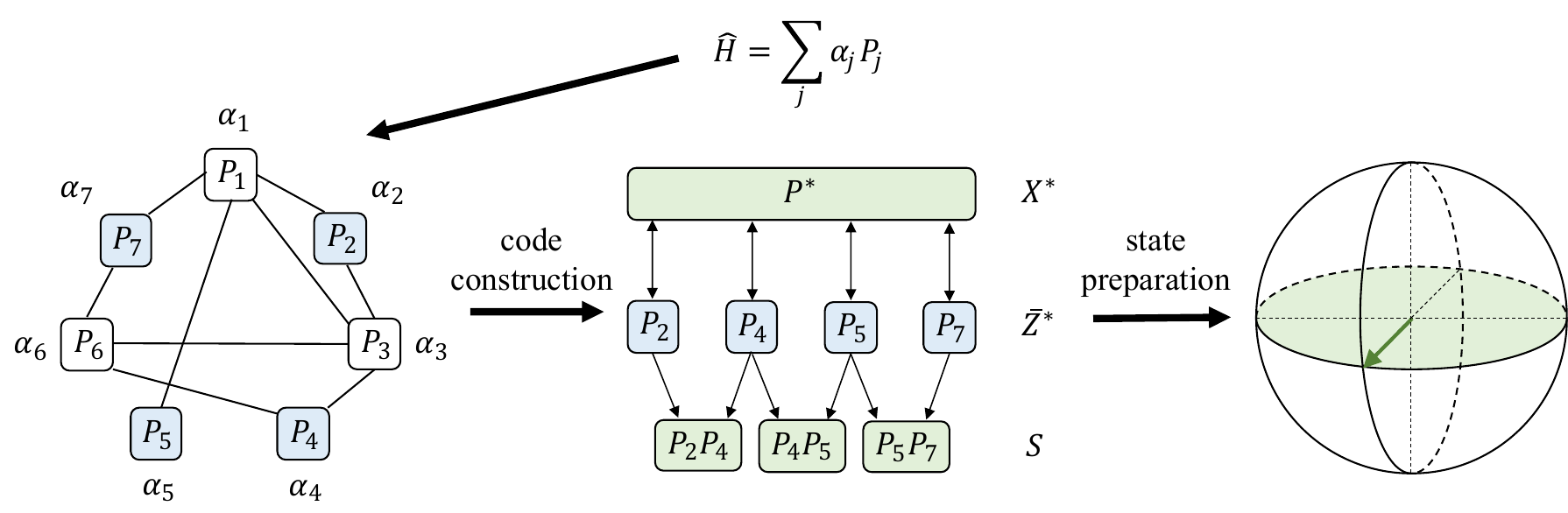}
\caption{\textbf{Hamiltonian-adapted asymmetric-code construction.}
Starting from a local Hamiltonian $\hat{H}=\sum_j \alpha_j P_j$, the Pauli components define a weighted anticommutation graph. A high-weight independent set, illustrated by the selected terms $P_2,P_4,P_5,P_7$, gives mutually commuting local signal components. Pairwise products of these components, such as $P_2P_4$, $P_4P_5$, and $P_5P_7$, are imposed as stabilizers. The individual $P_j$ remains a constant-weight physical representative of $\bar{Z}^*$. A Pauli operator $P^*$ that anticommutes with all selected representatives defines the conjugate logical operator $\bar{X}^*$. After preparing such a code state, one can go to the first step of Fig~\ref{fig:procedure}.}
\label{fig:construction}
\end{figure}

We first isolate the property required of the selected Pauli components. Recall that after absorbing signs into the Pauli operators, we write the relevant Pauli decomposition with $\alpha_j>0$. A subset $\mathcal{A}(\hat{H})\subseteq[M]$ is called an admissible signal family if it satisfies the following conditions:
\begin{enumerate}
\item[1.] The selected Pauli operators $\{P_j:j\in\mathcal{A}(\hat{H})\}$ commute pairwise;
\item[2.] The selected Pauli operators have no inconsistent or odd-parity identity relation. More explicitly, the product of any odd number of Pauli operators in $\{P_j:j\in\mathcal{A}(\hat{H})\}$ cannot be $\pm\id$, the product of any even number of Pauli operators in $\{P_j:j\in\mathcal{A}(\hat{H})\}$ cannot be $-\id$;
\item[3.] The selected components carry extensive signal weight, $\sum_{j\in\mathcal{A}(\hat{H})}\alpha_j=\Omega(n)$, which implies $|\mathcal{A}(\hat{H})|=\Theta(n)$.
\end{enumerate}
The pairwise commutation in condition 1, together with the even-parity part of condition 2, guarantees that the group generated by pairwise products of selected Pauli operators is a consistent stabilizer group, namely an Abelian Pauli group that does not contain $-\id$. The odd-parity part of condition 2 ensures that no individual selected Pauli operator is generated by these pairwise products; hence, the selected $P_j$'s remain nontrivial representatives of the same signal-aligned logical operator rather than becoming stabilizers themselves. Finally, condition 3 ensures that the selected family captures an extensive part of the Hamiltonian, which is needed for the coherent signal contribution to give Heisenberg-scaling QFI.

An admissible signal family can be founded by focusing on the weighted anticommutation graph of the Hamiltonian decomposition, as shown in Fig.~\ref{fig:construction}. The choice of an admissible signal family is not unique, and there are several natural ways to select one. A canonical choice is a maximum commuting Pauli set, denoted by $\mathrm{MCP}(\hat{H})$, defined as an admissible signal family maximizing the total coefficient weight,
\begin{equation}\label{eq:MCP}
\mathrm{MCP}(\hat{H})=\arg\max_{\mathcal{A}\ \text{admissible}}\sum_{j\in\mathcal{A}}\alpha_j .
\end{equation}
Another useful choice is a maximum disjoint Pauli set,
\begin{equation}\label{eq:MDP}
\mathrm{MDP}(\hat{H})=\arg\max_{\mathcal{A}\subseteq[M]:\,\mathrm{supp}(P_j)\cap \mathrm{supp}(P_{j'})=\emptyset,\,\forall j\neq j'\in \mathcal{A}}\sum_{j\in\mathcal{A}}\alpha_j .
\end{equation}
The disjointness condition makes $\mathrm{MDP}(\hat{H})$ automatically commuting and admissible. Since each Pauli component has weight at most $K$ and each qubit participates in at most $K$ Pauli components of the Hamiltonian, the overlap graph of the Pauli components has bounded degree. Thus, a standard greedy packing argument gives $\sum_{j\in\mathrm{MDP}(\hat{H})}\alpha_j=\Omega(n)$, which also gives a concrete example to ensure the existence of an admissible signal family for any $\hat{H}$. In the present subsection, the construction applies to any admissible signal family $\mathcal{A}(\hat{H})$. Taking $\mathcal{A}(\hat{H})=\mathrm{MCP}(\hat{H})$ gives the canonical asymmetric code. In the next subsection, we will take $\mathcal{A}(\hat{H})=\mathrm{MDP}(\hat{H})$ and use a modified low-density stabilizer layout to obtain strongly asymmetric QLDPC codes.

We now construct an asymmetric code $\mathcal{C}_{\hat{H}}$ based on an admissible signal family $\mathcal{A}(\hat{H})$. Choose a reference element $j_0\in\mathcal{A}(\hat{H})$, and define the preliminary stabilizer generator set and stabilizer group
\begin{equation}
\mathcal{S}_1=\left\{P_{j_0}P_j:\,j\in\mathcal{A}(\hat{H}),\,j\neq j_0\right\},\qquad \mathbb{S}_1=\langle\mathcal{S}_1\rangle.
\end{equation}
Since all operators in $\mathcal{A}(\hat{H})$ commute pairwise, the generators in $\mathcal{S}_1$ commute. Moreover, by admissibility, these generators define a consistent stabilizer group. Equivalently, every state $\ket{\psi}\in\mathcal{C}_{\hat{H}}$ satisfies $P_{j_1}P_{j_2}\ket{\psi}=\ket{\psi}$ for all $j_1$,$ j_2$ in $\mathcal{A}(\hat{H})$.

Each individual operator $P_j$ with $j\in\mathcal{A}(\hat{H})$ commutes with $\mathbb{S}_1$, but is not contained in $\mathbb{S}_1$. Furthermore, since $P_{j_1}P_{j_2}\in\mathcal{S}_1$, all selected $P_j$'s represent the same logical Pauli operator. We denote this logical operator by $\bar{Z}^*$. Then, we can choose another Pauli operator $P^*$ satisfying
\begin{equation}\label{eq:requirementS*}
\{P_j,P^*\}=0,\qquad \forall\, j\in\mathcal{A}(\hat{H}).
\end{equation}
Such a Pauli operator always exists, since a Pauli operator has $2n$ free parameters, but Eq.~\eqref{eq:requirementS*} gives at most $n$ independent consistent linear equations on them base on the second condition of the admissible signal family. Because $P^*$ anticommutes with each $P_j$, it commutes with every product $P_{j_1}P_{j_2}$ and therefore commutes with $\mathbb{S}_1$. We identify its logical operator with the conjugate logical operator $\bar{X}^*$.

If we stop here, the resulting code is asymmetric in the sense required for metrology. The proof is given in Supplementary Materials. The group $\mathbb{S}_1$ is the minimal stabilizer structure needed to define the signal-aligned logical direction and get an asymmetric code. We may further enlarge the stabilizer group by adding commuting Pauli generators, provided that the logical pair $\bar{Z}^*,\bar{X}^*$ is preserved. Concretely, any additional generator must commute with all selected representatives $P_j$, $j\in\mathcal{A}(\hat{H})$, and with $P^*$.

An important extension is the strongly asymmetric codes, which ensure that any other logical operators except for $\bar{Z}^*$ have non-constant weight. Starting from an empty set, $\mathcal{S}_2=\emptyset$, we iteratively add independent constant-weight Pauli operators that commute with every element of $\mathcal{S}_1\cup\mathcal{S}_2\cup{P^*}$. The procedure terminates when no further independent constant-weight Pauli operators satisfying these commutation relations can be added. We denote the resulting set by $\mathcal{S}_2^*$, and define $\mathcal{C}_{\hat{H}}^*$ as the stabilizer code generated by $\mathcal{S}_1\cup\mathcal{S}_2^*$.

The added generators preserve the logical operator $\bar{X}^*$ because they commute with $P^*$ by construction. They also preserve the signal-aligned logical operator $\bar{Z}^*$. Indeed, let $Q$ be any constant-weight Pauli operator commuting with $\mathcal{S}_1$. Since the stabilizers in $\mathcal{S}_1$ identify all selected operators $P_j$, $j\in\mathcal{A}(\hat{H})$, as representatives of the same logical operator, the commutation parity of $Q$ with $P_j$ is the same for all selected $P_j$'s. Thus, $Q$ either commutes with all selected $P_j$'s or anticommutes with all of them. The latter case is impossible for constant-weight $Q$: to anticommute with every selected local Pauli operator, the support of $Q$ must intersect the support of every $P_j\in\mathcal{A}(\hat{H})$, which requires weight $\Omega(|\mathcal{A}(\hat{H})|)=\Omega(n)$ under the bounded-degree locality assumptions. Therefore, every constant-weight operator added commutes with all selected representatives $P_j$, and $\bar{Z}^*$ remains well defined in the completed code. Moreover, the maximality of the construction removes all other constant-weight logical directions, making the code strongly asymmetric. Rigorous proof is given in Supplementary Material.

Finally, let $\ket{\psi}$ be a pure code state obtained by fixing the logical eigenvalue $\bar{X}^*\ket{\psi}=\ket{\psi}$, with any remaining logical degrees of freedom fixed arbitrarily. For every $j\in\mathcal{A}(\hat{H})$, Eq.~\eqref{eq:requirementS*} implies $\bra{\psi}P_j\ket{\psi}=0$, whereas the stabilizer relation gives $\bra{\psi}P_{j_1}P_{j_2}\ket{\psi}=1$ for all $j_1,j_2\in\mathcal{A}(\hat{H})$. Hence the selected Hamiltonian component $\hat{H}_{\rm MCP}=\sum_{j\in\mathcal{A}(\hat{H})}\alpha_jP_j$ acts coherently as a single logical signal, giving
\begin{equation}
\mathcal{F}(\ket{\psi},\hat{H}_{\mathcal{A}})=\left(\sum_{j\in\mathcal{A}(\hat{H})}\alpha_j\right)^2=\Omega(n^2).
\end{equation}
This explains why the construction restores Heisenberg-limited sensitivity: many local signal components become indistinguishable within the probe state and therefore add coherently. The mechanism applies to both the minimal asymmetric code and the completed strongly asymmetric code, since the added stabilizers preserve $\bar{Z}^*$ and $\bar{X}^*$. The full proof that the constructed code state achieves Heisenberg-limited QFI for the complete Hamiltonian $\hat{H}$ is given in the Supplementary Material.

\subsection{Construction of strongly asymmetric QLDPC codes}

The strongly asymmetric construction above removes all constant-weight logical directions except the signal-aligned one, but it does not by itself guarantee the full QLDPC condition, because constant check weight does not automatically imply bounded qubit participation. We now give an explicit low-density construction. The key is to use the freedom in choosing the signal-aligned Pauli family: instead of a maximum commuting Pauli set, we choose a maximum disjoint Pauli set, which captures an extensive amount of the Hamiltonian under the bounded-degree locality assumptions, while allowing the stabilizer generators to be chosen with both constant weight and bounded participation.

We use the definition of the maximum disjoint Pauli set of $\hat{H}$ in Eq.~\eqref{eq:MDP}. Write $\mathrm{MDP}(\hat{H})={j_1,\ldots,j_r}$ with $r=\Theta(n)$, and denote $B_{\ell}=\mathrm{supp}(P_{j_{\ell}})$. By construction, the supports $B_1,\ldots,B_r$ are mutually disjoint and each has size at most $K$.

We first construct the signal-aligned stabilizers. Instead of using a star-like generating set, we impose the pairwise identifications along a chain, $\left\{ P_{j_{\ell}}P_{j_{\ell+1}}:\ell=1,\ldots,r-1\right\}$. These generators commute because the selected Pauli operators have disjoint supports. They identify all $P_{j_{\ell}}$ as physical representatives of the same logical Pauli operator, which we denote by $\bar{Z}^*$. Moreover, every generator has weight at most $2K$, and each selected block $B_{\ell}$ participates in at most two such chain checks.

We next complete the code locally inside each selected support. For every $\ell$, choose a Pauli operator $P^*_{\ell}$ supported on $B_{\ell}$ such that $\{P^*_{\ell},P_{j_{\ell}}\}=0$. Such an operator always exists because $P_{j_{\ell}}$ is nontrivial on $B_{\ell}$. We then choose $|B_{\ell}|-1$ independent Pauli operators $Q_{\ell,1},\ldots,Q_{\ell,|B_{\ell}|-1}$ supported on $B_{\ell}$, mutually commuting, and commuting with both $P_{j_{\ell}}$ and $P^*_{\ell}$. Equivalently, these local checks define a one-logical-qubit stabilizer code on $B_{\ell}$, with logical operators represented by $P_{j_{\ell}}$ and $P^*_{\ell}$. This choice can always be made by completing the anticommuting pair $P_{j_{\ell}},P^*_{\ell}$ to a local symplectic basis on the qubits in $B_{\ell}$.

For qubits not contained in any selected support, we add single-qubit stabilizers, for example $Z_i$, to fix them. The contribution of these qubits can always be made non-negative by inverting the signs of the stabilizers, like from $Z_i$ to $-Z_i$. Therefore, they will not affect the scaling of the QFI. The full stabilizer group is then
\begin{equation}
\mathbb{S}^*=\left\langle P_{j_{\ell}}P_{j_{\ell+1}},,Q_{\ell,a},,Z_i:\ell=1,\ldots,r-1,\ a=1,\ldots,|B_{\ell}|-1,\ i\notin\bigcup_{\ell=1}^r B_{\ell}\right\rangle.
\end{equation}
The corresponding stabilizer code is denoted by $\mathcal{C}_{\hat{H}}^*$.

This construction satisfies the QLDPC condition. Every stabilizer generator has constant weight: the chain checks $P_{j_{\ell}}P_{j_{\ell+1}}$ have weight at most $2K$, the local checks $Q_{\ell,a}$ have weight at most $K$, and the unused-qubit checks have weight one. Moreover, each physical qubit participates in only $O(1)$ generators. Indeed, a qubit inside $B_{\ell}$ appears in at most $|B_{\ell}|-1=O(1)$ local checks and at most two chain checks, while a qubit outside all selected supports appears in one single-qubit check. Thus, $\mathcal{S}^*$ has a low-density generating set.

The logical structure is also explicit. Before the chain checks are imposed, each selected support $B_{\ell}$ carries one local logical qubit with logical operators $P_{j_{\ell}}$ and $P^*_{\ell}$. The chain checks $P_{j_{\ell}}P_{j_{\ell+1}}$ imposes a repetition-code constraint on the block logical qubits, leaving a single global logical qubit. The signal-aligned logical operator is represented by any selected Pauli operator, $\bar{Z}^*\sim P_{j_{\ell}}$, and a conjugate logical operator is represented by $\bar{X}^*=\prod_{\ell=1}^{r}P^*_{\ell}$. The operator $\bar{X}^*$ commutes with all chain checks because it anticommutes with both endpoints of each chain check, and it commutes with all local checks by construction.

One can also find that the code is strongly asymmetric. Since $\bar{Z}^*$ has a representative $P_{j_{\ell}}$ of weight at most $K$, the signal-aligned direction has constant effective weight. By contrast, any representative of a logical operator anticommuting with $\bar{Z}^*$ must anticommute with every physical representative $P_{j_{\ell}}$ of $\bar{Z}^*$. Because the supports $B_{\ell}$ are disjoint, such an operator must have nontrivial support on every selected block, and hence has weight at least $r=\Theta(n)$. Since the code encodes only one logical qubit, this means that $\bar{Z}^*$ is the unique logical direction with a constant-weight representative, while every logical operator independent of $\bar{Z}^*$ has effective weight $\Theta(n)$.

Finally, the metrological mechanism is the same as in the general asymmetric construction. The QFI argument does not rely on the particular choice $\mathcal{A}(\hat{H})=\mathrm{MDP}(\hat{H})$, but only on the admissibility of $\mathcal{A}(\hat{H})$ and the extensive-weight condition. The role of $\mathrm{MDP}(\hat{H})$ here is to give a general sufficient construction in which the stabilizer generators have both constant weight and bounded qubit participation. In practice, this choice need not be optimal for metrology: another admissible signal family $\mathcal{A}(\hat{H})$ may have a larger total coefficient weight and hence yield a larger QFI. Moreover, if the corresponding strongly asymmetric construction also admits a bounded-weight and bounded-participation generating set, the same construction gives a strongly asymmetric QLDPC code beyond the specific $\mathrm{MDP}(\hat{H})$ choice.

\subsection{Construction of concatenated asymmetric codes}

Concatenated asymmetric codes provide a tunable version of the asymmetric construction above. The codes constructed in the previous subsections keep the signal-aligned logical direction locally accessible, which is optimal for sensitivity but gives only constant effective distance in that direction. Here we increase the signal-direction distance by dividing the system into blocks, constructing an inner asymmetric code inside each block, and then coupling the block-level signal logical operators through an outer classical code. The block size $L$ controls the trade-off: larger blocks allow more local Hamiltonian components to add coherently and hence give larger QFI, whereas more blocks increase the distance of the signal-aligned logical direction.

We fix a block size $L$ and write $q=n/L$, assuming for simplicity that $L\mid n$. For general $L$, we take $q=\lfloor n/L\rfloor$ and pad or discard at most $O(L)$ qubits, which does not affect the asymptotic scaling. We consider $\sqrt{n}\leq L\leq n$. In the nontrivial concatenated regime $L=o(n)$, one has $q=\omega(1)$ and $q\leq L$. The endpoint $L=\Theta(n)$ recovers the fully asymmetric construction up to constant factors.

To localize the asymmetric construction inside blocks of size $L$, we use a restricted admissible signal family $\mathcal{A}_L(\hat{H})\subseteq[M]$, which is defined as an admissible signal family on $L$ rather than all the $n$ qubits. The definition of it can be obtained by simply replacing those $n$ in the definition of admissible signal family by $L$. The choice of an admissible signal family is also not unique, and $L$-restricted maximal commuting Pauli set and $L$-restricted maximal disjoint Pauli set still apply.

We now choose the blocks recursively. Set $R_0=[M]$. For $\ell=1,\ldots,q$, define
\begin{equation}
\mathcal{I}_{\ell}=\mathcal{A}_{L}(\hat{H}_{R_{\ell-1}}),\quad B_{\ell}=\bigcup_{j\in\mathcal{I}_{\ell}}\mathrm{supp}(P_j).
\end{equation}
By definition, $|B_{\ell}^{(0)}|\leq L$. If $|B_{\ell}^{(0)}|<L$, we enlarge it to a block $B_{\ell}$ of size $L$ by adding arbitrary qubits not used in the previously chosen blocks; otherwise, we set $B_{\ell}=B_{\ell}^{(0)}$. We then remove all Pauli terms whose support intersects the new block,
\begin{equation}
R_{\ell}=\{j\in R_{\ell-1}:\mathrm{supp}(P_j)\cap B_{\ell}=\emptyset\}.
\end{equation}
This procedure produces $q$ disjoint blocks $B_1,\ldots,B_q$, up to the irrelevant padding mentioned above.

For each block $B_{\ell}$, we construct an inner asymmetric code from the selected set $\mathcal{I}_{\ell}$. Specifically, for $j_1,j_2\in\mathcal{I}_{\ell}$, we impose $P_{j_1}P_{j_2}$ as stabilizers of the inner code. The individual Pauli operators $P_j$, $j\in\mathcal{I}_{\ell}$, then become different physical representatives of the same block logical operator, which we denote by $\bar{Z}_{\ell}^*$. By the same stabilizer-linear-algebra argument used above, one can choose a Pauli operator $P_{j_{\ell}}^*$ supported inside $B_{\ell}$ such that $\{P_{j_{\ell}}^*,P_j\}=0$ for all $j\in\mathcal{I}_{\ell}$. We identify the logical class of $P_{j_{\ell}}^*$ with the conjugate block logical operator $\bar{X}_{\ell}^*$.

We then complete each inner code at a tunable protection scale $W$, where $q\leq W\leq L$. This construction is the block analogue of the strongly asymmetric construction above, except that stabilizer generators of weight at most $W$ are allowed. Concretely, within each block, we add compatible independent Pauli generators of weight at most $W$ while preserving the logical pair $\bar{Z}_{\ell}^*,\bar{X}_{\ell}^*$, and continue until no further such generator can be added. The resulting inner stabilizer group, denoted by $\mathcal{S}^{\ell}$, is chosen so that $\bar{Z}_{\ell}^*$ remains the unique logical direction with constant-weight representatives, while every logical operator not generated by $\bar{Z}_{\ell}^*$ has minimum representative weight at least $W$.

It remains to concatenate the inner asymmetric codes using an outer classical code. Let $C_{\mathrm{cl}}\subseteq\mathbb{F}_2^q$ be a binary classical $[q,k,d_{\mathrm{cl}}]$ code with $d_{\mathrm{cl}}=\Theta(q)$, and let $H_{\mathrm{cl}}$ be a parity-check matrix of $C_{\mathrm{cl}}$. For each row $h\in\mathbb{F}_2^q$ of $H_{\mathrm{cl}}$, we add the outer stabilizer
\begin{equation}\label{eq:outerXstabilizers}
X(h)=\prod_{\ell:h_{\ell}=1}\bar{X}_{\ell}^*.
\end{equation}
We denote by $\mathcal{S}_{\mathrm{cl}}$ the stabilizer group generated by all such outer checks. The final concatenated asymmetric code is the stabilizer code with stabilizer group $\mathcal{S}_{\mathrm{con}}=\left\langle\mathcal{S}^1,\ldots,\mathcal{S}^q,\mathcal{S}_{\mathrm{cl}}\right\rangle$.

The role of the outer code is to convert the block-level signal representatives into logical operators whose block support must be a codeword of $C_{\mathrm{cl}}$. Thus, any nontrivial signal-aligned logical operator has support on at least $d_{\mathrm{cl}}=\Theta(q)$ blocks, while each block representative $\bar{Z}_{\ell}^*$ has a constant-weight physical representative. Since $W\geq q$, the signal-direction distance of the concatenated code is therefore $\Theta(q)=\Theta(n/L)$. Logical operators not generated by the block signal directions inherit the inner protection and have effective weight at least $\Omega(W)$, which can be chosen as large as $\Theta(L)$. Meanwhile, within each block, the selected Pauli components remain different representatives of the same block logical signal and hence contribute coherently to the QFI. As a result, suitable stabilizer sectors of the concatenated code contain probe states with QFI scaling as $\Theta(nL)$. The detailed distance and QFI proofs, including the stabilizer-sign choices, are given in the Supplementary Material.

In conclusion, the properties of different kinds of constructed codes are shown in the following table.

\begin{table*}[htbp]
\centering
\small
\setlength{\tabcolsep}{10pt}
\caption{Direction-resolved protection and sensitivity of asymmetric constructions. Here, $d_{\mathrm{sig}}$ represents the distance in the signal direction, $d_{\mathrm{conj}}$ represents the distance in the conjugate direction of the signal, and $d_{\mathrm{comp}}$ represents the distance in the complementary directions. $w$ represents the maximum stabilizer-generator weight of the code.}
\label{table:constructions}
\begin{tabular}{@{}cccccc@{}}
\toprule
Constructions
& $d_{\mathrm{sig}}$
& $d_{\mathrm{conj}}$
& $d_{\mathrm{comp}}$
& QFI $\mathcal{F}$
& $w$ \\
\midrule
General asymmetric codes
& $\Theta(1)$
& $\Theta(n)$
& $\Omega(1)$
& $\Theta(n^2)$
& $\Omega(1)$ \\
\addlinespace
Strongly asymmetric QLDPC codes
& $\Theta(1)$
& $\Theta(n)$
& $\omega(1)$
& $\Theta(n^2)$
& $O(1)$ \\
\addlinespace
Concatenated asymmetric code
& $\Theta(n/L)$
& $\Theta(L)$
& $\Omega(W)$, $O(L)$
& $\Theta(nL)$
& $O(L)$ \\
\bottomrule
\end{tabular}
\end{table*}

\clearpage
\onecolumngrid
\appendix

\begin{center}
	{\large \textbf{Supplementary Material}}
\end{center}

\makeatletter
\let\oldaddcontentsline\addcontentsline
\renewcommand{\addcontentsline}[3]{%
    \def\target{#1}%
    \def\toclabel{toc}%
    \ifx\target\toclabel
        \oldaddcontentsline{atoc}{atoc#2}{#3}%
    \else
        \oldaddcontentsline{#1}{#2}{#3}%
    \fi
}
    
\@starttoc{atoc}
\makeatother

\vspace{1cm}
    
\renewcommand{\thetheorem}{S\arabic{theorem}}
\renewcommand{\thelemma}{S\arabic{lemma}}
\renewcommand{\thedefinition}{S\arabic{definition}}
\renewcommand{\theproposition}{S\arabic{proposition}}
\renewcommand{\thecorollary}{S\arabic{corollary}}
\renewcommand{\theclaim}{S\arabic{claim}}
\renewcommand{\thepage}{S\arabic{page}}
\renewcommand{\thefigure}{S\arabic{figure}}
\renewcommand{\theHtheorem}{S\arabic{theorem}}
\renewcommand{\theHlemma}{S\arabic{lemma}}
\renewcommand{\theHdefinition}{S\arabic{definition}}
\renewcommand{\theHproposition}{S\arabic{proposition}}
\renewcommand{\theHcorollary}{S\arabic{corollary}}
\renewcommand{\theHclaim}{S\arabic{claim}}
\renewcommand{\theHfigure}{S\arabic{figure}}

\setcounter{theorem}{0}
\setcounter{lemma}{0}
\setcounter{equation}{0}
\setcounter{definition}{0}
\setcounter{proposition}{0}
\setcounter{claim}{0}
\setcounter{corollary}{0}
\setcounter{figure}{0}
\setcounter{page}{1}
\setcounter{section}{0}
\setcounter{equation}{0}

At the beginning, we collect the notation and assumptions used throughout the Supplementary Information. We consider an $n$-qubit $K$-local Hamiltonian with $K=O(1)$,
\begin{equation}\label{eq:HdecomPauliSI}
\hat{H}=\sum_{j=1}^{m}\hat{H}_j=\sum_{j=1}^{M}\alpha_j P_j,
\end{equation}
where each local term $\hat{H}_j$ acts nontrivially on at most $K$ qubits, satisfies $\|\hat{H}_j\|_{\infty}\leq 1$, and each qubit participates in $O(1)$ local terms. Thus $m=O(n)$ and $M\leq m4^K=O(n)$. The operators $P_j$ are Pauli operators of weight at most $K$. Identity components of the Hamiltonian are omitted, since they only generate a global phase and do not affect the QFI. By absorbing the signs of the coefficients into the Pauli operators, we take $\alpha_j\geq 0$ without loss of generality. Unless otherwise stated, asymptotic notation is taken in the limit $n\to\infty$, with $K$ constant fixed.

Throughout this work, we use the convention in which the QFI of a pure state is the variance of the sensing Hamiltonian,
\begin{equation}
\mathcal{F}(\ket{\psi},\hat{H})=\bra{\psi}\hat{H}^2\ket{\psi}-\left|\bra{\psi}\hat{H}\ket{\psi}\right|^2 .
\end{equation}
Equivalently, the QFI can be written as a sum of connected correlation functions between Hamiltonian terms. Expanding further in the Pauli basis gives
\begin{equation}\label{eq:QFIdecomPauliSI}
\begin{aligned}
\mathcal{F}(\ket{\psi},\hat{H})&=\sum_{j,j'=1}^m\left(\bra{\psi}\hat{H}_j\hat{H}_{j'}\ket{\psi}-\bra{\psi}\hat{H}_j\ket{\psi}\bra{\psi}\hat{H}_{j'}\ket{\psi}\right)\\
&=\sum_{j_1,j_2=1}^M\alpha_{j_1}\alpha_{j_2}\left(\bra{\psi}P_{j_1}P_{j_2}\ket{\psi}-\bra{\psi}P_{j_1}\ket{\psi}\bra{\psi}P_{j_2}\ket{\psi}\right).
\end{aligned}
\end{equation}
This decomposition will be used repeatedly below. It shows that superlinear QFI requires many nonvanishing connected correlations among local Pauli components of the Hamiltonian. Conversely, if the code structure forces such connected correlations to vanish except for $O(n)$ pairs of local terms, then the QFI is bounded by the standard quantum limit scaling, $\mathcal{F}=O(n)$.

We also fix the code-theoretic notation used in the following proofs. For a stabilizer code with stabilizer group $\mathcal{S}$, $\mathrm{wt}(P)$ denotes the Pauli weight of $P$, and the effective weight of a logical Pauli operator $\bar{O}$ is
\begin{equation}
\mathrm{wt}_{\mathrm{eff}}(\bar{O})=\min_{S\in\mathcal{S}}\mathrm{wt}(S\bar{O}) .
\end{equation}
The code distance is the minimum effective weight over all nontrivial logical Pauli operators.

\section{Proof of the trade-off results}\label{appsec:proofobstruction}

In the following subsections, we prove the protection-sensitivity trade-off relations for non-degenerate codes, QLDPC codes and generalized Shor codes, respectively. For non-degenerate codes and QLDPC codes, we obtain an even stronger limitation: under growing-distance protection, every code state has at most $\mathcal{F}=O(n)$ for any local sensing Hamiltonian satisfying the assumptions above. Generalized Shor codes evade this strict standard-quantum-limit bound by using degeneracy, but they still obey a quantitative trade-off between code distance and metrological sensitivity.

\subsection{QFI bound for non-degenerate codes}\label{appsec:proofnondege}

We first show that non-degenerate QEC codes with growing distance cannot support superlinear QFI under local Hamiltonian evolution. The reason is that non-degeneracy makes distinct correctable local Pauli errors orthogonal, thereby eliminating the connected correlations between spatially separated local Hamiltonian terms.

\begin{lemma}\label{lemma:nondegenerate}
Let $\hat{H}=\sum_{j=1}^m \hat{H}_j$ be a $K$-local Hamiltonian, and let $\ket{\psi}$ be a code state of a non-degenerate QEC code with distance $d$. If $d\geq 2K+1$, then the QFI satisfies
\begin{equation}\label{eq:boundQEC}
\mathcal{F}(\ket{\psi},\hat{H}) \leq mK^2 .
\end{equation}
In particular, since $m=\Theta(n)$ and $K=O(1)$, the QFI scales at most linearly with the system size and therefore cannot surpass the standard quantum limit.
\end{lemma}

The proof follows by combining the Knill-Laflamme conditions with the covariance decomposition of the QFI. Since all $K$-local Pauli components are correctable errors, non-degeneracy implies that two distinct such Pauli errors have zero covariance on any code state; hence only overlapping local Hamiltonian terms can contribute.

\begin{proof}
Let $\Pi$ be the projector onto the code space. Since $d\geq 2K+1$, all Pauli errors of weight at most $K$ are correctable. Moreover, because the code is non-degenerate, distinct correctable Pauli errors have orthogonal error spaces. Equivalently, after fixing a phase convention for Pauli operators, the Knill-Laflamme conditions imply
\begin{equation}\label{eq:KLnondegenerate}
\Pi P^\dagger P'\Pi=\delta_{P,P'}\Pi
\end{equation}
for all Pauli operators $P,P'$ of weight at most $K$. In particular, for every non-identity Pauli operator $P$ of weight at most $K$, taking $P'=I$ gives $\Pi P\Pi=0$.

We first show that two local Hamiltonian terms with disjoint supports have zero covariance on any code state. Suppose $\mathrm{supp}(\hat{H}_j)\cap\mathrm{supp}(\hat{H}_{j'})=\emptyset$. Expanding both terms in the Pauli basis, every non-identity Pauli operator $P$ appearing in $\hat{H}_j$ and every non-identity Pauli operator $P'$ appearing in $\hat{H}_{j'}$ has weight at most $K$. Since the supports are disjoint, $P\neq P'$. Therefore, Eq.~\eqref{eq:KLnondegenerate} gives $\bra{\psi}P\ket{\psi}=0$, $\bra{\psi}P'\ket{\psi}=0$, and $\bra{\psi}PP'\ket{\psi}=0$. By linearity, we obtain $\bra{\psi}\hat{H}_j\hat{H}_{j'}\ket{\psi}-\bra{\psi}\hat{H}_j\ket{\psi}\bra{\psi}\hat{H}_{j'}\ket{\psi}=0$. Thus, in the term-wise covariance decomposition of the QFI, a nonzero contribution can arise only from pairs of local Hamiltonian terms whose supports overlap.

It remains to count such pairs. For a fixed local term $\hat{H}_j$, its support contains at most $K$ qubits. By the bounded-degree assumption, each of these qubits participates in at most $K$ local Hamiltonian terms. Thusm there are at most $K^2$ indices $j'$ such that $\mathrm{supp}(\hat{H}_j)\cap\mathrm{supp}(\hat{H}_{j'})\neq\emptyset$. For each overlapping pair, the covariance is bounded by
\begin{equation}
\left|\bra{\psi}\hat{H}_j\hat{H}_{j'}\ket{\psi}-\bra{\psi}\hat{H}_j\ket{\psi}\bra{\psi}\hat{H}_{j'}\ket{\psi}\right|\leq\sqrt{\mathrm{Var}_{\psi}(\hat{H}_j)\mathrm{Var}_{\psi}(\hat{H}_{j'})}\leq 1,
\end{equation}
where we used $\|\hat{H}_j\|_{\infty}\leq 1$ and $\|\hat{H}_{j'}\|_{\infty}\leq 1$. Summing over all ordered pairs of local Hamiltonian terms gives
\begin{equation}
\mathcal{F}(\ket{\psi},\hat{H})\leq \sum_{j=1}^m\sum_{j':\,\mathrm{supp}(\hat{H}_j)\cap\mathrm{supp}(\hat{H}_{j'})\neq\emptyset}1\leq mK^2,
\end{equation}
which proves the lemma.
\end{proof}

\subsection{QFI bound for QLDPC codes}\label{appsec:proofQLDPC}

We next prove an analogous standard-quantum-limit bound for QLDPC stabilizer codes. In contrast to non-degenerate codes, a degenerate stabilizer code may allow nonzero correlations between distinct local Pauli errors when their product is a stabilizer. The QLDPC condition, however, makes such stabilizer-induced correlations local in the Tanner graph, so only a constant number of Hamiltonian terms can be correlated with any fixed local term.

\begin{lemma}\label{lemma:qldpc}
Let $\hat{H}=\sum_{j=1}^{m}\hat{H}_j$ be a $K$-local Hamiltonian, and let $\ket{\psi}$ be a code state of an $n$-qubit QLDPC stabilizer code with distance $d$. If $d\geq 2K+1$, then the QFI satisfies
\begin{equation}
\mathcal{F}(\ket{\psi},\hat{H})=O(n).
\end{equation}
Thus, under local Hamiltonian evolution, QLDPC code states with growing distance cannot surpass the standard quantum limit.
\end{lemma}

The proof uses two facts. First, because $d\geq 2K+1$, any Pauli operator of weight at most $2K$ has nonzero expectation on a code state only if it is a stabilizer up to a phase. Second, in a QLDPC code, a low-weight stabilizer cannot create correlations between local Pauli operators that are far apart in the Tanner graph.

\begin{proof}
Let $\mathcal{S}$ be the stabilizer group, and let $S_1,\ldots,S_r$ be a QLDPC generating set. We denote by $s=O(1)$ a uniform upper bound on both the weight of each stabilizer generator and the number of stabilizer generators acting on any physical qubit. Let $G=(V^Q,V^S,E)$ be the Tanner graph associated with this generating set: vertices in $V^Q$ correspond to physical qubits, vertices in $V^S$ correspond to stabilizer generators, and an edge connects a qubit vertex to a stabilizer vertex whenever the corresponding generator acts nontrivially on that qubit. For two subsets of vertices $A,B\subseteq V^Q$, we write $\mathrm{dist}(A,B)$ for their graph distance in $G$.

We first characterize when a Pauli covariance term in Eq.~\eqref{eq:QFIdecomPauliSI} can be nonzero. Let $P_{j_1}$ and $P_{j_2}$ be Pauli operators of weight at most $K$. Since $d\geq 2K+1$, the product $P_{j_1}P_{j_2}$ has weight smaller than $d$. Therefore, if $P_{j_1}P_{j_2}$ has nonzero expectation on a code state, it must be a stabilizer up to a phase; otherwise it is either detectable and has zero expectation, or it would be a nontrivial logical operator of weight below the code distance. Similarly, each individual $P_{j_a}$ has nonzero expectation only if it is a stabilizer up to a phase. If either $P_{j_1}$ or $P_{j_2}$ is a stabilizer, then the covariance vanishes because multiplication by a stabilizer only fixes the sign of the expectation value. Hence a nonzero covariance term can occur only when $P_{j_1}P_{j_2}$ is a stabilizer up to a phase while neither $P_{j_1}$ nor $P_{j_2}$ is itself a stabilizer.

We now show that such a nonzero covariance is impossible when the two Pauli operators are separated in the Tanner graph. Let $V^Q_{j_1},V^Q_{j_2}\subseteq V^Q$ denote the supports of $P_{j_1}$ and $P_{j_2}$. Suppose that $\mathrm{dist}(V^Q_{j_1},V^Q_{j_2})>2$. Then no stabilizer generator can intersect both supports. Assume, for contradiction, that $P_{j_1}P_{j_2}$ is a stabilizer up to a phase while neither $P_{j_1}$ nor $P_{j_2}$ is a stabilizer. Since $P_{j_1}P_{j_2}\in\mathcal{S}$ up to a phase, it commutes with every stabilizer generator $S_\ell$. If $S_\ell$ intersects the support of $P_{j_1}$, then it is disjoint from the support of $P_{j_2}$, and hence $[P_{j_2},S_\ell]=0$. From $[P_{j_1}P_{j_2},S_\ell]=0$, it follows that $[P_{j_1},S_\ell]=0$. If $S_\ell$ does not intersect the support of $P_{j_1}$, then it commutes with $P_{j_1}$ trivially. Thus $P_{j_1}$ commutes with all stabilizer generators and hence lies in the normalizer of $\mathcal{S}$. Since $\mathrm{wt}(P_{j_1})\leq K<d$, $P_{j_1}$ cannot be a nontrivial logical operator; it must be a stabilizer, contradicting the assumption. Therefore, Pauli covariance terms can be nonzero only when the corresponding Pauli supports have Tanner distance at most $2$.

It remains to count how many local Hamiltonian terms can be correlated with a fixed one. For a fixed term $\hat{H}_j$, its support contains at most $K$ qubits. In the Tanner graph, the number of qubit vertices within distance $2$ of this support is $O(Ks^2)$. By the bounded-degree assumption on the Hamiltonian, each physical qubit participates in at most $K$ local Hamiltonian terms. Hence only $O(K^2s^2)$ local Hamiltonian terms $\hat{H}_{j'}$ can contain Pauli components whose supports lie within Tanner distance $2$ of a Pauli component of $\hat{H}_j$. For all other $j'$, every Pauli covariance term between $\hat{H}_j$ and $\hat{H}_{j'}$ vanishes, and therefore $\bra{\psi}\hat{H}_j\hat{H}_{j'}\ket{\psi}-\bra{\psi}\hat{H}_j\ket{\psi}\bra{\psi}\hat{H}_{j'}\ket{\psi}=0$.

For the remaining $O(K^2s^2)$ terms, the covariance is bounded by $1$. Summing over all ordered pairs of local Hamiltonian terms gives
\begin{equation}
\mathcal{F}(\ket{\psi},\hat{H})\leq \sum_{j=1}^{m}O(K^2s^2)=O(m)=O(n),
\end{equation}
because $K=O(1)$, $s=O(1)$, and $m=\Theta(n)$. This proves the lemma.
\end{proof}

\subsection{Distance-QFI trade-off in generalized Shor codes}

We now consider degenerate codes for which the linear QFI bound proved above need not hold. A simple example is the $[[L^2,1,L]]$ Shor code. Let $n=L^2$ and consider the local Hamiltonian $\hat{H}=\sum_{a=1}^{L}\sum_{b=1}^{L} Z_{a,b}$, where $Z_{a,b}$ acts on the $b$-th qubit in the $a$-th block. Take the code state $\ket{\psi}=\left(\frac{\ket{0}^{\otimes L}+\ket{1}^{\otimes L}}{\sqrt{2}}\right)^{\otimes L}$. For each block, the operator $\sum_{b=1}^{L} Z_{a,b}$ has variance $L^2$ on the corresponding GHZ state, while different blocks are uncorrelated. Hence, the QFI $\mathcal{F}(\ket{\psi},\hat{H})=L\cdot L^2=L^3=n^{3/2}$. At the same time, the code distance is $d=L=n^{1/2}$. Thus, the Shor code has non-constant distance while still achieving superlinear QFI scaling. The reason is precisely its degeneracy: products of local $Z$ operators within the same block can be low-weight stabilizers, allowing the corresponding local signal terms to add coherently.

To isolate this mechanism, we consider generalized Shor codes.

\begin{definition}[Generalized Shor code]
Let $C_1=[n_1,k_1,d_1]$ and $C_2=[n_2,k_2,d_2]$ be two binary linear codes, with parity-check matrices $H_1$ and $H_2$, and let $G_1$ be a generator matrix of $C_1$. The generalized Shor code associated with $(C_1,C_2)$ is a CSS code on $n_1n_2$ physical qubits. We arrange the qubits into $n_2$ blocks $B_1,\cdots,B_{n_2}$, each containing $n_1$ qubits. In the CSS convention used here, the $Z$-type stabilizers are generated by $H_1\otimes I_{n_2}$, while the $X$-type stabilizers are generated by $G_1\otimes H_2$, and hence couple different blocks through codewords of $C_1$. The resulting quantum code has parameters $[[n_1n_2,k_1k_2,\min(d_1,d_2)]]$.
\end{definition}

The ordinary $[[L^2,1,L]]$ Shor code is recovered by taking both $C_1$ and $C_2$ to be length-$L$ repetition codes. Generalized Shor codes therefore provide a useful testbed for quantifying the protection-sensitivity trade-off in degenerate codes. On the one hand, their local $Z$-type stabilizers can create nonzero covariances among signal terms supported in the same block, as in the Shor-code example above. On the other hand, the same block structure restricts how many local Hamiltonian terms can be coherently correlated with any fixed local term once the code distance is required to grow. Thus, although generalized Shor codes can evade the linear-QFI limitation for non-degenerate and QLDPC codes, their metrological enhancement is still constrained by the amount of protection they provide. The following lemma makes this constraint quantitative.

\begin{lemma}\label{lemma:Shor}
Let $\hat{H}=\sum_{j=1}^{m}\hat{H}_j$ be an $n$-qubit $K$-local Hamiltonian. Suppose that $\ket{\psi}$ is a code state of an $[[n_1n_2,k_1k_2,\min(d_1,d_2)]]$ generalized Shor code, where $n=n_1n_2$ and $d=\min(d_1,d_2)\geq 2K+1$. Then the QFI obeys
\begin{equation}\label{eq:Shor}
d\cdot \mathcal{F}(\ket{\psi},\hat{H}) \leq K^2mn .
\end{equation}
\end{lemma}

The proof again uses the covariance decomposition of the QFI. Since $P_{j_1}P_{j_2}$ has weight at most $2K<d$, a nonzero Pauli covariance can only arise when $P_{j_1}P_{j_2}$ is a stabilizer. For generalized Shor codes, the only such low-weight stabilizers are $Z$-type stabilizers inside the blocks, so a fixed local Hamiltonian term can be correlated only with terms touching the same blocks.

\begin{proof}
Let $\mathcal{S}$ be the stabilizer group of the generalized Shor code. We first characterize which Pauli covariance terms in Eq.~\eqref{eq:QFIdecomPauliSI} can be nonzero. For any Pauli operators $P_{j_1}$ and $P_{j_2}$ appearing in the Pauli decomposition of $\hat{H}$, both of them have weight at most $K$. Therefore, $P_{j_1}P_{j_2}$ cannot be a nontrivial logical operator. As a result, a nonzero Pauli covariance term can occur only when $P_{j_1}P_{j_2}$ is a stabilizer up to a sign while neither $P_{j_1}$ nor $P_{j_2}$ is itself a stabilizer.

We next use the specific stabilizer structure of generalized Shor codes. The $Z$-type stabilizer group generated by $H_1\otimes I_{n_2}$ factors block by block. In contrast, every stabilizer with a nontrivial $X$ component contains the $X$-support of a nonzero codeword of $C_1$, and therefore has $X$-support of weight at least $d_1$. Multiplication by $Z$-type stabilizers cannot reduce this $X$-support. Since $d_1\geq d>2K$, no Pauli operator of weight at most $2K$ can be a stabilizer with nontrivial $X$ component. Consequently, if $P_{j_1}P_{j_2}$ contributes nontrivially to the covariance, then $P_{j_1}P_{j_2}$ must be a $Z$-type stabilizer.

We now show that such a covariance can occur only when the supports of $P_{j_1}$ and $P_{j_2}$ intersect a common block. Suppose, to the contrary, that no block contains qubits from both supports. Since $P_{j_1}P_{j_2}$ is a $Z$-type stabilizer and the $Z$-type stabilizer group factors over blocks, the restriction of $P_{j_1}P_{j_2}$ to each block is itself a $Z$-type stabilizer on that block. Because the supports of $P_{j_1}$ and $P_{j_2}$ occupy disjoint sets of blocks, this implies that $P_{j_1}$ and $P_{j_2}$ are each products of blockwise $Z$-type stabilizers, and hence are themselves stabilizers up to signs. This contradicts the condition for a nonzero connected covariance. Therefore, any pair contributing to the QFI must have supports intersecting at least one common block.

It remains to count the possible Hamiltonian terms that can be correlated with a fixed one. Fix a local term $\hat{H}_j$. Its support contains at most $K$ qubits and therefore intersects at most $K$ blocks. These blocks contain at most $Kn_1$ physical qubits in total. By the previous paragraph, any local term $\hat{H}_{j'}$ that can have nonzero covariance with $\hat{H}_j$ must act on at least one qubit inside these blocks. Since each qubit participates in at most $K$ Hamiltonian terms, the number of such choices of $j'$ is at most $K^2n_1$.

For each potentially contributing pair, the covariance is bounded by $1$. Summing over all ordered pairs of local Hamiltonian terms gives
\begin{equation}
\mathcal{F}(\ket{\psi},\hat{H})\leq mK^2n_1.
\end{equation}
Finally, since $d=\min(d_1,d_2)\leq d_2\leq n_2$, we obtain
\begin{equation}
d\cdot \mathcal{F}(\ket{\psi},\hat{H})\leq n_2\cdot mK^2n_1=K^2mn,
\end{equation}
which proves the lemma.
\end{proof}

For bounded-degree local Hamiltonians with $m=\Theta(n)$ and $K=O(1)$, Lemma~\ref{lemma:Shor} gives the scaling form
\begin{equation}
d\cdot \mathcal{F}(\ket{\psi},\hat{H})=O(n^2).
\end{equation}
Thus, generalized Shor codes can use degeneracy to achieve superlinear QFI, but only by paying a proportional cost in code distance.

\section{Distance and QFI of asymmetric codes}

In this section, we prove the distance properties and QFI scaling of the constructed asymmetric codes, including general/strongly asymmetric codes, strongly asymmetric QLDPC codes, and concatenated asymmetric codes.

\subsection{Distance and QFI of asymmetric QEC codes}

In this subsection, we prove the logical-distance properties and the Heisenberg-limited QFI scaling of the asymmetric codes used in the main text. For the general asymmetric construction, the result is the existence of a code state with Heisenberg-limited QFI. For strongly asymmetric constructions, including the strongly asymmetric QLDPC construction, the result is stronger: after fixing the auxiliary stabilizer signs, every code state in the $\bar{X}^*=+1$ sector has Heisenberg-limited QFI.

We first verify the asymmetric logical structure. Since $\bar{Z}^*$ has a representative $P_j\in\mathcal{A}(\hat{H})$ of weight at most $K$, we have $\mathrm{wt}_{\mathrm{eff}}(\bar{Z}^*)\leq K=O(1)$. By contrast, every representative of $\bar{X}^*$ must anticommute with every selected Pauli operator $P_j\in\mathcal{A}(\hat{H})$. Since each selected $P_j$ is local and each physical qubit participates in only $O(1)$ selected terms, a Pauli operator of weight $w$ can anticommute with at most $O(w)$ of them. Therefore, $\mathrm{wt}_{\mathrm{eff}}(\bar{X}^*)=\Omega(|\mathcal{A}(\hat{H})|)=\Omega(n)$. Since the constructed representative $P^*$ has weight $O(n)$, we obtain $\mathrm{wt}_{\mathrm{eff}}(\bar{X}^*)=\Theta(n)$. Thus, the signal-aligned logical direction $\bar{Z}^*$ remains locally accessible, while the conjugate logical direction $\bar{X}^*$ has macroscopic effective weight.

For the strongly asymmetric construction, we further show that $\bar{Z}^*$ is the only logical direction with a constant-weight representative. In the construction used here, additional stabilizer generators are chosen to commute with $\mathcal{S}_1$, with $P^*$, and with one fixed signal representative $P_{j_0}\in\mathcal{A}(\hat{H})$; because every $P_j\in\mathcal{A}(\hat{H})$ differs from $P_{j_0}$ by an element of $\mathcal{S}_1$, this is equivalent to preserving both logical operators $\bar{Z}^*$ and $\bar{X}^*$. Now consider any logical Pauli operator $O$ of constant weight. Since $O$ commutes with $\mathcal{S}_1$, it either commutes with all $P_j\in\mathcal{A}(\hat{H})$ or anticommutes with all of them. In the latter case, bounded-degree locality implies $\mathrm{wt}(O)=\Omega(n)$, contradicting the assumption that $O$ has constant weight. Hence, any constant-weight logical operator must commute with all selected $P_j$. If such an $O$ also commutes with $P^*$ and is independent of the stabilizer group, then it satisfies the commutation conditions for the construction and would have been added as an auxiliary stabilizer, contradicting maximality. If instead $O$ anticommutes with $P^*$, then $OP_{j_0}$ has constant weight and commutes with $P^*$ as well as with all selected $P_j$. By the previous argument, $OP_{j_0}$ must be trivial up to stabilizers; hence $O$ is equivalent to $P_{j_0}$ and represents the signal-aligned logical operator $\bar{Z}^*$. Therefore, every constant-weight logical operator is equivalent to $\bar{Z}^*$, while all logical operators independent of $\bar{Z}^*$ have non-constant effective weight. This proves strong asymmetry.

It remains to prove the QFI scaling. We first consider the general asymmetric code. Let $\ket{\psi}\in\mathcal{C}_{\hat{H}}$ be a stabilizer state satisfying $\bar{X}^*\ket{\psi}=\ket{\psi}$. Let $\mathcal{S}_1'=\mathcal{S}_1\cup\{\bar{X}^*\}$, with all signs fixed to be positive, and let $\mathcal{S}_2'$ denote the remaining stabilizer generators needed to specify $\ket{\psi}$. We analyze Eq.~\eqref{eq:QFIdecomPauliSI} by separating the covariance terms according to whether $\pm P_{j_1}P_{j_2}$ lies in the group generated by $\mathcal{S}_1'$ or outside that group.

If either $\pm P_{j_1}$ or $\pm P_{j_2}$ is itself a stabilizer of $\ket{\psi}$, then the corresponding covariance term vanishes. Hence, a nonzero covariance term can occur only when neither individual operator is a stabilizer up to sign, while $P_{j_1}P_{j_2}$ is a stabilizer up to sign. In that case, the sign of the covariance contribution is determined by the stabilizer sign of $P_{j_1}P_{j_2}$.

We first treat all terms for which $P_{j_1}P_{j_2}$ is not contained in the group generated by $\mathcal{S}_1'$ up to sign. The total contribution of these terms can always be made nonnegative by an appropriate choice of the signs of the generators in $\mathcal{S}_2'$. To see this, define the signed projector associated with a set of stabilizer generators $\mathcal{S}$ by
\begin{equation}
\label{eq:signedprojector}
\Pi_{\vec{b}}(\mathcal{S})\equiv \prod_{S\in\mathcal{S}}\frac{\id+(-1)^{b_S}S}{2},
\end{equation}
where $\vec{b}=(b_S)_{S\in\mathcal{S}}$ is a binary vector. For each sign vector $\vec{b}$, define the corresponding stabilizer state by
\begin{equation}
\ketbra{\psi_{\vec{b}}}=\Pi_{\vec{0}}(\mathcal{S}_1')\Pi_{\vec{b}}(\mathcal{S}_2').
\end{equation}
Averaging over all choices of $\vec{b}$ gives
\begin{equation}
\begin{aligned}
\sum_{\vec{b}}\bra{\psi_{\vec{b}}}P_{j_1}P_{j_2}\ket{\psi_{\vec{b}}}&=\tr{P_{j_1}P_{j_2}\sum_{\vec{b}}\ketbra{\psi_{\vec{b}}}}\\
&=\tr{P_{j_1}P_{j_2}\Pi_{\vec{0}}(\mathcal{S}_1')\sum_{\vec{b}}\prod_{S\in\mathcal{S}_2'}\frac{\id+(-1)^{b_S}S}{2}}\\
&=\tr{P_{j_1}P_{j_2}\Pi_{\vec{0}}(\mathcal{S}_1')}=0.
\end{aligned}
\end{equation}
The last equality follows from Pauli orthogonality, because $P_{j_1}P_{j_2}$ is not contained in the group generated by $\mathcal{S}_1'$ up to sign. Summing this identity over all covariance terms of the above type shows that their average total contribution over the sign choices of $\mathcal{S}_2'$ is zero. Therefore, at least one choice of the sign vector $\vec{b}$ makes their total contribution nonnegative. We fix such a sign choice in the following.

We now consider the positive contributions enforced by $\mathcal{S}_1'$. For any $P_{j_1},P_{j_2}\in\mathcal{A}(\hat{H})$, the product $P_{j_1}P_{j_2}$ belongs to $\mathcal{S}_1$ with positive sign by construction. Moreover, each selected $P_j$ anticommutes with $\bar{X}^*$, and hence $\bra{\psi}P_j\ket{\psi}=0$. Therefore, the selected Pauli components contribute
\begin{equation}
\label{eq:positivelowerbound_re}
\sum_{P_{j_1},P_{j_2}\in\mathcal{A}(\hat{H})}\alpha_{j_1}\alpha_{j_2}=\left(\sum_{P_j\in\mathcal{A}(\hat{H})}\alpha_j\right)^2=\Theta(n^2)
\end{equation}
to the QFI, by the definition of $\mathcal{A}(\hat{H})$.

We next bound the possible negative contributions from terms satisfying $-P_{j_1}P_{j_2}\in\langle\mathcal{S}_1'\rangle$. Such a term cannot involve the logical factor $\bar{X}^*$, because $P_{j_1}P_{j_2}$ has weight at most $2K=O(1)$, whereas every representative of $\bar{X}^*$ has effective weight $\Theta(n)$. Thus, negative contributions of this type can only come from low-weight elements generated by $\mathcal{S}_1$, namely from local products of selected Pauli operators. Products involving more than two selected operators are handled by the same bounded-degree counting below and change only the constant prefactor, so it suffices to count the possible local products of the form $P_{j_1'}P_{j_2'}$ with $P_{j_1'},P_{j_2'}\in\mathcal{A}(\hat{H})$.

Introduce the support graph of the Pauli decomposition of $\hat{H}$: the vertices are physical qubits, and two vertices are connected if there exists a Pauli term $P_j$ that acts nontrivially on both of them. This graph has bounded degree, denoted $O(K^2)$. If there is no edge connecting a vertex in $\supp(P_{j_1'})$ to a vertex in $\supp(P_{j_2'})$, then a relation $-P_{j_1}P_{j_2}=P_{j_1'}P_{j_2'}$ cannot occur except for trivial relabelings that do not produce a negative covariance contribution. Hence, only selected pairs with overlapping constant-size neighborhoods need to be considered. For each fixed $P_{j_1'}$, there are at most $O(K^4)$ choices of $P_{j_2'}$ whose supports lie in the relevant neighborhood. Since the total number of Pauli terms is $M=O(n)$, the number of such selected pairs is at most $O(MK^4)=O(n)$. For each fixed product $P_{j_1'}P_{j_2'}$, the number of Hamiltonian-Pauli pairs $(j_1,j_2)$ satisfying $-P_{j_1}P_{j_2}=P_{j_1'}P_{j_2'}$ is bounded by a constant depending only on $K$ and the local degree; in particular, it may be bounded by $2^{\mathrm{wt}(P_{j_1'}P_{j_2'})}\leq 2^{2K}$ up to such constant local-degree factors. Therefore, the total number of negatively contributing pairs is bounded by
\begin{equation}
O(MK^4 2^{2K})=O(n),
\end{equation}
and, since the coefficients $\alpha_j$ are bounded, their total contribution is at most $O(n)$.

Combining the positive contribution from Eq.~\eqref{eq:positivelowerbound_re}, the nonnegative contribution obtained by the sign choice of $\mathcal{S}_2'$, and the $O(n)$ bound on possible negative terms, we obtain
\begin{equation}
\mathcal{F}(\ket{\psi},\hat{H})\geq \Theta(n^2)-O(n)=\Omega(n^2).
\end{equation}
The opposite bound $\mathcal{F}(\ket{\psi},\hat{H})\leq \|\hat{H}\|_{\infty}^2=O(n^2)$ follows from the extensivity of $\hat{H}$. Therefore,
\begin{equation}
\mathcal{F}(\ket{\psi},\hat{H})=\Theta(n^2).
\end{equation}
Thus, the general asymmetric code $\mathcal{C}_{\hat{H}}$ contains a code state with Heisenberg-limited QFI.

Finally, consider the strongly asymmetric construction, including the strongly asymmetric QLDPC construction. The proof above uses only the stabilizers in $\mathcal{S}_1$ and the logical constraint $\bar{X}^*\ket{\psi}=\ket{\psi}$ to obtain the extensive positive contribution. Additional stabilizers can only affect the remaining local covariance terms through their signs, and these signs remain free parameters of the code construction. We choose them once using the sign-averaging argument above. After this choice, the same lower bound holds for every code state in the $\bar{X}^*=+1$ sector. Indeed, by strong asymmetry, no $O(1)$-weight Pauli operator can represent any logical direction independent of $\bar{Z}^*$; hence the expectation values of all local Pauli operators and of all products of two local Pauli operators are independent of the remaining logical state, except for the fixed stabilizer signs and the fixed $\bar{X}^*$ eigenvalue. Therefore, the selected signal terms always have zero one-point expectation values and positive pairwise correlations, while all possible negative local contributions are still bounded by $O(n)$. Consequently, every code state of the strongly asymmetric code satisfying $\bar{X}^*\ket{\psi}=\ket{\psi}$ obeys $\mathcal{F}(\ket{\psi},\hat{H})=\Theta(n^2)$. The same argument applies to the explicit strongly asymmetric QLDPC codes, since they realize the same strongly asymmetric logical structure while keeping the stabilizer generators low-density.

\subsection{Distance and QFI of concatenated asymmetric codes}

We now prove the distance and QFI scaling of the concatenated asymmetric codes. We focus on the nontrivial concatenated regime $q=n/L=\omega(1)$, with $q\leq W\leq L$ and $d_{\mathrm{cl}}=\Theta(q)$. The endpoint $q=O(1)$ reduces, up to constant factors, to the fully asymmetric construction discussed above.

We first analyze the distance of the code. Before adding the outer stabilizers $\mathcal{S}_{\mathrm{cl}}$, the code is the tensor product of the $q$ inner asymmetric codes. By the inner-code completion, any logical operator with representative weight less than $W$ must be generated, up to inner stabilizers, by the block signal logical operators $\bar{Z}_{\ell}^*$. Hence any such logical operator can be written as
\begin{equation}
Z(c)=\prod_{\ell:c_{\ell}=1}\bar{Z}_{\ell}^*,\quad c\in\mathbb{F}_2^q.
\end{equation}
After adding the outer stabilizers, $Z(c)$ commutes with every outer check
\begin{equation}
X(h)=\prod_{\ell:h_{\ell}=1}\bar{X}_{\ell}^*
\end{equation}
if and only if $h\cdot c=0$ for every row $h$ of $H_{\mathrm{cl}}$. This condition is equivalent to $c\in C_{\mathrm{cl}}$. Therefore, every nontrivial logical operator generated by the signal-aligned block logicals has support on at least $d_{\mathrm{cl}}$ blocks. Since each active block contributes at least constant physical weight, every such operator has weight at least $\Omega(d_{\mathrm{cl}})=\Omega(q)=\Omega(n/L)$. Conversely, a minimum-weight codeword of $C_{\mathrm{cl}}$ gives a logical operator of weight $O(d_{\mathrm{cl}})$, because each $\bar{Z}_{\ell}^*$ has a constant-weight representative. Thus the signal-aligned logical distance is $\Theta(d_{\mathrm{cl}})$.

All logical operators not generated by the block signal logicals inherit the inner protection. Indeed, if such an operator had a representative weight smaller than $W$, then its restriction to each inner block would have to be generated by the corresponding $\bar{Z}_{\ell}^*$, contradicting the assumption that the operator is not generated by the block signal logicals. Hence all logical operators outside the signal-aligned sector have weight at least $\Omega(W)$. Combining the two cases, the distance of the concatenated code is
\begin{equation}\label{eq:concatenatedDistance}
d_{\mathrm{con}}=\Theta\!\left(\min\{W,d_{\mathrm{cl}}\}\right)=\Theta(q)=\Theta(n/L),
\end{equation}
where we used $W\geq q$ and $d_{\mathrm{cl}}=\Theta(q)$.

In particular, if the outer code is the repetition code $[q,1,q]$ and $W=\omega(q)$, then the unique minimum-weight logical operator is, up to stabilizers, the signal logical operator
\begin{equation}
\bar{Z}_{\mathrm{sig}}=\prod_{\ell=1}^q \bar{Z}_{\ell}^*,
\end{equation}
with weight $\Theta(q)=\Theta(n/L)$. Every other nontrivial logical operator has weight at least $\Omega(W)$. Thus, the concatenated code is strongly asymmetric: the signal direction has tunable distance $\Theta(n/L)$, whereas all the complementary logical directions are protected up to the inner scale $W$, which can be chosen as large as $\Theta(L)$.

We now prove the QFI scaling. Let
\begin{equation}
w_{\ell}=\sum_{j\in\mathcal{I}_{\ell}}\alpha_j
\end{equation}
be the total coefficient weight selected in block $B_{\ell}$. The definition of the restricted admissible signal family gives $\sum_{\ell=1}^q w_{\ell}^2=\Omega(nL)$. Consider any pure state $\ket{\psi}$ in the concatenated code space after fixing the stabilizer signs as specified below. For every $j_1,j_2\in\mathcal{I}_{\ell}$, the product $P_{j_1}P_{j_2}$ is an inner stabilizer with positive sign. On the other hand, each individual $P_j$, $j\in\mathcal{I}_{\ell}$, anticommutes with at least one outer stabilizer. To see this, note that if coordinate $\ell$ did not participate in any parity check of $H_{\mathrm{cl}}$, then the unit vector supported on $\ell$ would belong to $C_{\mathrm{cl}}$, contradicting $d_{\mathrm{cl}}>1$. Therefore, for every $j\in\mathcal{I}_{\ell}$ there exists a row $h$ of $H_{\mathrm{cl}}$ with $h_{\ell}=1$, so that $P_j$ anticommutes with $X(h)$. Since $\ket{\psi}$ is stabilized by all outer checks, this implies $\langle \psi|P_j|\psi\rangle=0$ for all $j\in\mathcal{I}_{\ell}$. At the same time, the positive inner stabilizer signs give $\langle \psi|P_{j_1}P_{j_2}|\psi\rangle=1$ for all $j_1,j_2\in\mathcal{I}_{\ell}$. Thus, the selected Pauli pairs inside block $B_{\ell}$ contribute
\begin{equation}
\sum_{j_1,j_2\in\mathcal{I}_{\ell}}\alpha_{j_1}\alpha_{j_2}=w_{\ell}^2
\end{equation}
to Eq.~\eqref{eq:QFIdecomPauliSI}. Summing over all blocks gives a positive contribution $\sum_{\ell=1}^q w_{\ell}^2=\Omega(nL)$.

It remains to control possible negative contributions. Since the final code has distance $\Theta(q)=\omega(1)$, any operator $P_{j_1}P_{j_2}$ of constant weight cannot be a nontrivial logical operator for sufficiently large $n$. Therefore, a nonzero covariance term can arise only when $P_{j_1}P_{j_2}$ is a stabilizer up to a sign. Moreover, a constant-weight operator cannot be generated nontrivially from the outer stabilizers in $\mathcal{S}_{\mathrm{cl}}$. Indeed, any nontrivial product of the outer checks acts as a non-signal logical operator $\bar{X}_{\ell}^*$ in at least one inner block, and therefore has physical representative weight at least $\Omega(W)=\omega(1)$ even after multiplication by inner stabilizers. Hence all constant-weight stabilizer products relevant to Eq.~\eqref{eq:QFIdecomPauliSI} are generated within the inner block codes.

The signs of the additional completion stabilizers inside the blocks can be chosen by the same sign-averaging argument used for the strongly asymmetric codes, so that their total contribution is nonnegative. The only remaining negative terms are those satisfying $-P_{j_1}P_{j_2}=P_{j'_1}P_{j'_2}$ for some selected pair $j'_1,j'_2$ inside an inner block. This counting is purely local and is unchanged by the concatenation. As in the previous subsection, for each fixed product $P_{j'_1}P_{j'_2}$, the number of pairs $(j_1,j_2)$ satisfying the above identity is bounded by a constant depending only on $K$; in particular, it is at most $2^{\mathrm{wt}(P_{j'_1}P_{j'_2})}\leq 2^{2K}$ up to bounded-degree factors. The number of relevant selected products is at most $O(MK^4)=O(n)$. Therefore, the total number of negative contributions is bounded by
\begin{equation}
O(MK^4 2^{2K})=O(n),
\end{equation}
and, since the coefficients $\alpha_j$ are bounded, their total magnitude is also $O(n)$.

Combining the positive contribution, the nonnegative contribution from the sign choice of the completion stabilizers, and the $O(n)$ bound on possible negative terms, we obtain
\begin{equation}\label{eq:concatenatedQFI_lower}
\mathcal{F}(\ket{\psi},\hat{H})\geq\Omega(nL)-O(n)=\Omega(nL),
\end{equation}
where the last equality uses $L=\omega(1)$.

A matching upper bound follows from the same block-locality structure. As argued above, a constant-weight Pauli product contributing to Eq.~\eqref{eq:QFIdecomPauliSI} cannot involve the outer stabilizers nontrivially; it must be generated within the inner blocks. Hence, a fixed Hamiltonian Pauli component $P_j$ can have nonzero connected covariance only with Pauli components whose supports touch one of the $O(1)$ blocks intersecting $\supp(P_j)$. These blocks contain $O(L)$ physical qubits in total, and bounded-degree locality implies that only $O(L)$ Hamiltonian Pauli components can act there. Since there are $M=O(n)$ Pauli components in the Hamiltonian decomposition and each covariance term has bounded magnitude, we have $\mathcal{F}(\ket{\psi},\hat{H})=O(nL)$. Together, we have
\begin{equation}\label{eq:concatenatedQFI}
\mathcal{F}(\ket{\psi},\hat{H})=\Theta(nL).
\end{equation}

Thus, the concatenated asymmetric code realizes a tunable protection-sensitivity tradeoff. The signal-direction distance is $\Theta(n/L)$, the complementary logical directions have effective distance at least $\Omega(W)$, and the QFI scales as $\Theta(nL)$. Choosing $W=\Theta(L)$ gives the strongest complementary protection for a fixed block size $L$, while varying $L$ tunes the interpolation between the fully asymmetric Heisenberg-scaling construction and codes with larger signal-direction distance.

\section{Probe state preparation in quantum metrology}

In this section, we discuss the preparation of the probe states and the corresponding preparation lower bound. We first specify the adaptive circuit model used throughout this section. A $\kappa$-local gate is a channel of the form $\mathcal{E}=\mathcal{E}_A\otimes I_{\bar A}$ with $|A|\leq \kappa$, where $\mathcal{E}_A$ preserves the Hilbert-space dimension of subsystem $A$ and $I_{\bar A}$ is the identity channel on the remaining qubits. A mid-circuit measurement is modeled as a quantum instrument acting on at most $\kappa$ qubits; after the classical outcome is recorded, the measured qubits may be discarded or reset. An adaptive circuit prepares an output state $\rho$ from a product initial state, which we write compactly as $\rho=\mathcal{E}'_t\circ\cdots\circ\mathcal{E}'_2\circ\mathcal{E}'_1(\ketbra{\mathbf{0}})$. Here each $\mathcal{E}'_\tau$ denotes the averaged channel induced by one layer of disjoint $\kappa$-local gates and mid-circuit measurements, where the operations in layer $\tau$ may depend on all previous measurement outcomes. Within the same layer, the quantum supports of different gates and measurements are disjoint. The time cost is the number of layers $t$, with both unitary layers and measurement layers counted, and the space cost is the number of ancillary qubits used during the preparation. In our explicit preparation circuits below, measured ancillary qubits are discarded and never reused.

\subsection{Preparation of concatenated asymmetric code states}

In this subsection, we discuss the preparation of the probe code states. We begin with a general lemma showing that the joint measurement of commuting Pauli operators can be implemented by a constant-depth adaptive Clifford circuit, at the cost of ancillary qubits proportional to the total check weight.

\begin{lemma}
\label{lemma:commuting_pauli_measurement}
Let $P_1,\ldots,P_t$ be mutually commuting nontrivial Hermitian Pauli operators on $n$ qubits. Denote $w_i=\mathrm{wt}(P_i)$ and $W_{\mathrm{tot}}=\sum_{i=1}^t w_i$. The joint measurement of $P_1,\ldots,P_t$ can be realized by a constant-depth adaptive Clifford circuit using $O(W_{\mathrm{tot}}+n)$ ancillary qubits. More precisely, for every input state $\rho$, the circuit outputs a bit string $\mathbf{x}\in\{0,1\}^t$ with probability $\tr{\Pi_{\mathbf{x}}\rho}$ and implements the measurement instrument
\begin{equation}
\rho\longmapsto \Pi_{\mathbf{x}}\rho\Pi_{\mathbf{x}},\qquad \Pi_{\mathbf{x}}=\prod_{i=1}^t\frac{\id+(-1)^{x_i}P_i}{2}.
\end{equation}
Equivalently, conditioned on an outcome with nonzero probability, the post-measurement state is $\Pi_{\mathbf{x}}\rho\Pi_{\mathbf{x}}/\tr{\Pi_{\mathbf{x}}\rho}$.
\end{lemma}

\begin{proof}
The proof has two steps. We first construct the standard Pauli-measurement circuit, which has size $O(W_{\mathrm{tot}})$. We then compress this Clifford circuit to constant depth using a measurement-based, or equivalently gate-teleportation, implementation of Clifford circuits.

For each Pauli operator $P_i$, introduce a syndrome ancilla $a_i$ initialized in $\ket{+}$. Write $P_i=\bigotimes_{j\in\mathrm{supp}(P_i)}P_{i,j}$, where identities outside $\mathrm{supp}(P_i)$ are omitted. For each $j\in\mathrm{supp}(P_i)$, apply the controlled-Pauli gate $CP_{i,a_i,j}=\ketbra{0}_{a_i}\otimes \id+\ketbra{1}_{a_i}\otimes P_{i,j}$. This is a two-qubit Clifford gate. The product $V_i=\prod_{j\in\mathrm{supp}(P_i)}CP_{i,a_i,j}$ equals the controlled-$P_i$ operation with control $a_i$. Since the Pauli operators $P_i$ commute, the controlled operations $V_i$ also commute. After applying all $V_i$, measuring each syndrome ancilla $a_i$ in the $X$ basis implements the projective measurement of $P_i$. Hence the full circuit implements the joint projective measurement with projectors $\Pi_{\mathbf{x}}$. The number of two-qubit Clifford gates is $W_{\mathrm{tot}}$, because one controlled-Pauli gate is used for each qubit in the support of each measured Pauli operator, and the number of syndrome ancillas is $t\leq W_{\mathrm{tot}}$.

It remains to reduce the depth. We use the standard measurement-based implementation of Clifford circuits~\cite{Gottesman1999Demonstrating,Raussendorf2002Computational}. A Clifford circuit on $n+t$ wires with $O(W_{\mathrm{tot}})$ two-qubit Clifford gates and a final layer of Pauli measurements can be converted into a graph-state measurement pattern with $O(W_{\mathrm{tot}}+n+t)=O(W_{\mathrm{tot}}+n)$ qubits and bounded graph degree. Such a bounded-degree graph state can be prepared in constant depth: color the graph edges using $O(1)$ colors and apply all controlled-$Z$ gates of one color in parallel. After the graph state is prepared, the input state is teleported into the graph by Bell measurements, and all internal graph qubits are measured in Pauli bases. Since the simulated circuit is Clifford, the measurement bases are fixed Pauli bases and do not depend on previous quantum measurement outcomes. The teleportation outcomes only produce Pauli byproducts, which can be handled by classical postprocessing of the reported syndrome bits and, if desired, by one final layer of classically controlled one-qubit Pauli corrections on the output register. Therefore, all measurements can occur in a constant number of measurement layers, which completes the proof.
\end{proof}

Lemma~\ref{lemma:commuting_pauli_measurement} immediately gives a preparation procedure for the strongly asymmetric code states constructed above. First initialize a product eigenstate of $\bar{X}^*=P^*$ with eigenvalue $+1$. This is possible because $P^*$ is a Pauli string, so one can initialize each qubit in an eigenstate of the corresponding single-qubit Pauli operator. Then measure the stabilizer generators of the strongly asymmetric code with the desired signs. The measurement outcomes may differ from the target signs, but this can be corrected without changing the eigenvalue of $\bar{X}^*$. Applying this correction, or equivalently tracking it in the Pauli frame, prepares a code state satisfying $\bar{X}^*\ket{\psi}=\ket{\psi}$ and the desired stabilizer signs. For the strongly asymmetric construction, the total stabilizer-check weight is $O(n)$, and therefore the preparation uses a constant-depth adaptive Clifford circuit with $O(n)$ ancillary qubits.

We now discuss the preparation of concatenated asymmetric code states. The procedure has two steps. First, we prepare a state satisfying the outer stabilizers. This step does not require measuring the outer code. For each block $B_{\ell}$, choose a product state $\ket{\phi_{\ell}}$ that is a $+1$ eigenstate of the block Pauli operator $\bar{X}_{\ell}^*$, $\bar{X}_{\ell}^*\ket{\phi_{\ell}}=\ket{\phi_{\ell}}$. This is possible because $\bar{X}_{\ell}^*$ is a Pauli operator supported inside $B_{\ell}$. Define $\ket{\phi_0}=\bigotimes_{\ell=1}^q\ket{\phi_{\ell}}$. Then every outer stabilizer $X(h)=\prod_{\ell:h_{\ell}=1}\bar{X}_{\ell}^*$ has eigenvalue $+1$ on $\ket{\phi_0}$. Thus, the outer stabilizers are already satisfied at the beginning of the preparation.

Second, we measure a generating set of the inner stabilizers. This generating set includes, in each block, the signal-identification stabilizers, for example $P_{j_{\ell}^0}P_j$ for $j\in\mathcal{I}_{\ell}\setminus\{j_{\ell}^0\}$ with a fixed reference element $j_{\ell}^0\in\mathcal{I}_{\ell}$, together with the additional completion stabilizers constructed inside the block. The full inner stabilizer group then contains all products $P_{j_1}P_{j_2}$ with $j_1,j_2\in\mathcal{I}_{\ell}$, even though only a generating set needs to be measured. We choose the target signs of these inner generators as in the QFI proof: the signal-identification stabilizers have positive sign, while the auxiliary completion signs are chosen so that the remaining local covariance contribution is nonnegative.

By Lemma~\ref{lemma:commuting_pauli_measurement}, the joint measurement of the inner stabilizer generators can be implemented by a constant-depth adaptive Clifford circuit. There are $O(n)$ such generators, and each has weight at most $W$ in the concatenated construction. Hence the total check weight is $O(nW)$, and the preparation uses $O(nW)$ ancillary qubits.

It remains to verify that the correction step can be performed without disturbing the outer stabilizers. All inner stabilizer generators commute with every $\bar{X}_{\ell}^*$ by construction. Therefore, measuring the inner stabilizers preserves the eigenvalues of all $\bar{X}_{\ell}^*$, and hence preserves all outer stabilizers $X(h)$. After the measurement, the observed inner signs may differ from the target signs. As above, standard stabilizer linear algebra gives Pauli corrections that flip the required inner stabilizer signs while commuting with all $\bar{X}_{\ell}^*$. Applying these corrections, or tracking them in the Pauli frame, fixes the desired inner stabilizer signs and leaves the eigenvalue of every $\bar{X}_{\ell}^*$ unchanged. Consequently, the resulting state $\ket{\psi_{\mathrm{con}}}$ is stabilized by all inner stabilizers and all outer stabilizers with the desired signs, and therefore belongs to the concatenated code space $\mathcal{C}_{\mathrm{con}}$.

The $O(nW)$ ancillary-qubit count is a direct constructive upper bound, not a lower bound. The preparation lower bound proved in Lemma~\ref{lemma:LRE} below implies that any constant-depth adaptive preparation of a state with $\mathcal{F}(\rho,\hat{H})=\Omega(nL)$ must use at least $\Omega(L)$ effective ancillary qubits. Thus, for concatenated asymmetric code states, the known preparation cost lies between the lower bound $\Omega(L)$ and the direct stabilizer-measurement upper bound $O(nW)$. Closing this gap would require a more optimized preparation procedure than measuring all inner stabilizer generators in parallel through the generic Clifford-compression construction.

\subsection{Preparation bound for probe states}\label{appsec:proofLRE}

In this subsection, we prove a general preparation lower bound for metrologically useful probe states. The result is independent of the asymmetric-code constructions above: any state with Heisenberg-limited QFI must contain long-range correlations, and such correlations cannot be generated by a shallow adaptive circuit unless the circuit has sufficiently large effective space resources.

\begin{lemma}[Preparation bound for metrological probe states]
\label{lemma:LRE}
Let $\hat{H}=\sum_{j=1}^m \hat{H}_j$ be a $K$-local Hamiltonian. Suppose an $n$-qubit state $\rho$ is prepared from a product state by an adaptive circuit of depth $t$ using $\kappa$-qubit gates and $n_a$ ancillary qubits. Then
\begin{equation}\label{eq:LRE_bound}
\mathcal{F}(\rho,\hat{H}) \leq mK^2 (n_a+1) g(\kappa,2t),
\end{equation}
where $g(\kappa,2t)=\kappa^{2t}$ for all-to-all connectivity, and $g(\kappa,2t)=(4(\kappa-1)t+1)^r$ for an $r$-dimensional grid in which each gate acts on a $\kappa\times\kappa\times\cdots\times\kappa$ hypercube. In particular, since $m=\Theta(n)$ and $K=O(1)$, achieving Heisenberg-limited scaling $\mathcal{F}(\rho,\hat{H})=\Theta(n^2)$ requires
\begin{equation}
\label{eq:LRE_requirement}
(n_a+1)g(\kappa,2t)=\Omega(n).
\end{equation}
\end{lemma}

\begin{proof}
The proof has two steps. We first reduce the QFI to connected two-point correlation functions of local Hamiltonian terms. We then use a lightcone bound for adaptive circuits to show that only a limited number of these connected correlations can be nonzero.

Let $\rho=\sum_i \lambda_i \ketbra{\psi_i}$ be any pure-state decomposition of $\rho$. By the convexity of the QFI and by the pure-state variance formula, we have
\begin{equation}\label{eq:mixeddecom}
\begin{aligned}
\mathcal{F}(\rho,\hat{H})&\leq\sum_i\lambda_i\mathcal{F}(\ket{\psi_i},\hat{H})\\
&=\sum_i\lambda_i\left(\tr{\ketbra{\psi_i}\hat{H}^2}-\left(\tr{\ketbra{\psi_i}\hat{H}}\right)^2\right)\\
&\leq \tr{\rho \hat{H}^2}-\left(\tr{\rho \hat{H}}\right)^2\\
&=\sum_{j,j'=1}^m\left(\tr{\rho \hat{H}_j\hat{H}_{j'}}-\tr{\rho \hat{H}_j}\tr{\rho \hat{H}_{j'}}\right).
\end{aligned}
\end{equation}
Thus, the QFI is upper bounded by the sum of connected correlations among the local Hamiltonian terms.

To bound these correlations, we use a lightcone argument from Ref.~\cite{liu2025state}. The intuition is that in a circuit of bounded depth, local operators can influence only a limited effective neighborhood. As the following lemma shows, even with adaptivity, the region that can remain correlated with a given local observable grows only by a controlled factor depending on the depth and the available ancillas. Outside this enlarged lightcone, connected correlations vanish exactly.

\begin{lemma}[Ref.~\cite{liu2025state}]\label{lemma:preparecorrelation}
Suppose a state $\rho$ can be prepared by an adaptive circuit with depth $t$ and $n_a$ ancillas. For any set of qubits $A$, there exists a larger set $A'$ with $|A'|\leq (n_a+1) g(\kappa,2t)\cdot|A|$ such that, for any set of qubits $B$ with $B\cap A'=\emptyset$,
\begin{equation}
\tr{\rho O_1O_2}-\tr{\rho O_1}\tr{\rho O_2}=0,\qquad\forall\,\mathrm{supp}(O_1)=A,\ \mathrm{supp}(O_2)=B.
\end{equation}
Here, $g(\kappa,2t)=\kappa^{2t}$ for all-to-all connectivity and $g(\kappa,2t)=(4(\kappa-1)t+1)^r$ for an $r$-dimensional grid in which each gate acts on a $\kappa\times\kappa\times\cdots\times\kappa$ hypercube.
\end{lemma}

We apply the lemma with $A=\supp(\hat{H}_j)$. Since each $\hat{H}_j$ acts on at most $K$ qubits, the corresponding enlarged set satisfies $|A'|\leq (n_a+1) g(\kappa,2t)\cdot K$. A term $\hat{H}_{j'}$ can contribute to Eq.~\eqref{eq:mixeddecom} only if its support intersects $A'$, because otherwise the connected correlation vanishes by Lemma~\ref{lemma:preparecorrelation}. As each qubit participates in at most $K$ Hamiltonian terms, the number of indices $j'$ that can contribute for fixed $j$ is at most $K|A'|\leq (n_a+1) g(\kappa,2t)\cdot K^2$. Moreover, $\|\hat{H}_j\|_\infty\leq 1$ implies that each individual correlation is bounded in absolute value by $1$. Summing over all $j$ yields
\begin{equation}
\mathcal{F}(\rho,\hat{H}) \leq mK^2(n_a+1)g(\kappa,2t).
\end{equation}
This completes the proof.
\end{proof}

Lemma~\ref{lemma:LRE} shows that Heisenberg-limited metrological scaling requires extensive effective preparation resources. For constant-depth adaptive circuits, $g(\kappa,2t)=O(1)$, and therefore Eq.~\eqref{eq:LRE_requirement} requires $n_a=\Omega(n)$. Our strongly asymmetric QLDPC code states achieve $\mathcal{F}(\rho,\hat{H})=\Theta(n^2)$ while being preparable by constant-depth adaptive Clifford circuits using $O(n)$ ancillary qubits. Hence their preparation cost saturates the bound up to constant factors.


Finally, in the absence of ancillary qubits, Lemma~\ref{lemma:LRE} reduces to the usual lightcone obstruction for shallow non-adaptive state preparation. For all-to-all two-qubit circuits, Heisenberg-limited QFI requires $2^{2t}=\Omega(n)$, and hence $t=\Omega(\log n)$. For an $r$-dimensional local grid, it requires $(4(\kappa-1)t+1)^r=\Omega(n)$, and hence $t=\Omega(n^{1/r})$ up to constant factors. Therefore, Heisenberg-limited probe states cannot be generated by shallow circuits with sublinear effective spacetime resources.

\end{document}